\documentclass[floatfix,preprintnumbers,amssymb]{revtex4}

\usepackage{graphicx}

\usepackage{epsfig,psfrag,amsmath,amssymb,float,psfrag}
\newcommand{\be}{\begin{equation}} 

\newcommand{\ee}{\end{equation}}

\newcommand{\bea}{\begin{eqnarray}} 

\newcommand{\eea}{\end{eqnarray}}

\newcommand{\bc}{\begin{center}} 

\newcommand{\ec}{\end{center}} 

\begin{document}

\title{Critical phenomena and quantum phase transition in long range
  Heisenberg antiferromagnetic chains} 

\author{Nicolas Laflorencie, Ian Affleck and Mona Berciu}

\affiliation{Department of Physics \& Astronomy, University of British
Columbia, Vancouver, B.C., Canada, V6T 1Z1}

\begin{abstract}
Antiferromagnetic Hamiltonians with short-range, non-frustrating
interactions are well-known to exhibit long range magnetic order in
dimensions, $d\geq 2$ but exhibit only quasi long range order, with
power law decay of correlations, in $d=1$ (for half-integer spin).  On
the other hand, non-frustrating long range interactions can induce
long range order in $d=1$. We study Hamiltonians in which the long
range interactions have an adjustable amplitude $\lambda$, as well as
an adjustable power-law $1/|x|^\alpha$, using a combination of quantum
Monte Carlo and analytic methods: spin-wave, large-$N$ non-linear
$\sigma$ model, and renormalization group methods.  We map out the
phase diagram in the $\lambda$-$\alpha$ plane and study the nature of
the critical line separating the phases with long range and quasi long
range order. We find that this corresponds to a novel line of critical
points with continuously varying critical exponents and a dynamical
exponent, $z<1$.
\end{abstract}

\maketitle

\section{Introduction}

The ground-state (GS) of the nearest neighbor antiferromagnetic (AF)
Heisenberg model on a bipartite lattice:
\begin{equation} 
{\cal{H}}=\sum_{\langle i,j\rangle}\vec S_i\cdot \vec S_j,
\end{equation}
is generally expected to have long range order (LRO):
\begin{equation} 
\label{Neel}
\langle \vec S_{0}\cdot \vec S_r\rangle \to \pm m^{2}_{\rm{AF}},\ \
  (r\to \infty),
\end{equation} 
for any spin magnitude, $S$ and any dimension $d\geq
 2$~\cite{Ian94}. On the other hand, in dimension $d=1$, the behavior
 depends on whether $S$ is integer or
 half-integer~\cite{Haldane83-87}.  In the half-integer case the
 spin-spin correlation function
\begin{equation} 
\langle\vec S_0\cdot \vec S_r\rangle\propto \frac{(-1)^{r}{\sqrt{\ln
r}}}{r},
 \label{corr}
\end{equation}
 is expected~\cite{Giamarchi89}, characteristic of a {\it quasi}-long
 range order (QLRO).  (In the integer spin, Haldane gap case,
 correlations decay exponentially.)  This behavior in $d=1$ for
 half-integer $S$ is believed to be universal, not depending on the
 magnitude of $S$ nor on the details of the Hamiltonian as long as it
 is short range and not too frustrating. Long range interactions
 (i.e. power-law decaying with the relative distance between
 interacting moments) can be introduced in spin models either for some
 experimental reasons like dipolar or RKKY interactions or simply
 because of some theoretical relevance. For instance a famous example
 is the Haldane-Shastry model~\cite{HS88} with AF frustrating $1/r^2$
 interaction which exhibits an exact RVB GS. Another
 theoretical interest comes from the possibility to interpolate
 between discrete dimensions by tuning continuously the exponent that
 governs the decay of the interaction with the distance.  Indeed, the
 possibility for true LRO to occur in $d=1$ with long range
 interactions has motivated many studies during the last
 decades~\cite{MW66,68-69,Fisher72,Kosterlitz76,Brezin76,Cardy81,Frolich82,Luijten01,Bruno01,Yusuf04} 
 and is the subject of the present paper. While a long standing debate
 about the critical behavior of the Ising model in $d$ dimensions with
 long range ferromagnetic interaction decaying like $r^{-d-\sigma}$
 has been quite intense during the last thirty
 years~\cite{68-69,Frolich82,Luijten01}, the N-vector model
 has also been a subject of interest for many
 authors~\cite{Fisher72,Kosterlitz76,Brezin76,Cardy81}.  Concerning
 the Heisenberg model with long range interaction $\sim r^{-\alpha}$,
 the seminal paper of Mermin and Wagner~\cite{MW66} proving the
 absence of LRO at finite temperature $T$ in $d\le 2$ for $\alpha>d+2$
 has been recently reconsidered by Bruno~\cite{Bruno01} who gave
 stronger conditions for the absence of spontaneous magnetic order at
 $T>0$ in $d\le 2$. For instance, he proved that the AF non
 frustrating one dimensional model \be {\cal{H}} = \sum_i\left[ \vec
 S_i\cdot \vec S_{i+1} -\lambda \sum_{j=2}^\infty (-1)^{j}{\vec
 S_i\cdot \vec S_{i+j} \over j^{\alpha}}\right].
\label{QLRO}
\ee does not have N\'eel order for any temperature $T>0$ if $\alpha
\geq 2$.
Actually, much less
is known about the $T=0$ case, except the work of Parreira {\it et
al.}~\cite{Parreira97} where the authors signaled the existence of the
bound $\alpha=3$ over which $T=0$ LRO is ruled out~\cite{note1}. A
particular case, the model (\ref{QLRO}) with $\lambda=1$ was recently
analyzed at both $T=0$ and $T>0$ in Ref.~\cite{Yusuf04}, using the
lowest order spin-wave (SW) approximation, expected to be valid for
large enough $S$ and small enough $\alpha$. There it was shown that
the SW dispersion relation takes the {\it sublinear} form, at low
$k$:
\be
\label{SL}
\omega (k) \propto |k|^{(\alpha -1)/2}, \ee for $\alpha
<3$. Consequently the quantum $1/S$ reduction of the order parameter:
\be \Delta m_q \propto \int {dk\over \omega (k)},\ee is finite for any
$\alpha <3$. By requiring that $\Delta m_q < S$, a consistency
condition on the SW approximation, it is concluded that LRO occurs for
any $S$ at sufficiently small $\alpha$.  (However, such an estimate is
presumably only reliable for $S\gg1$.) After correcting a numerical
error in Ref.~\cite{Yusuf04}, the SW
prediction for the $S=1/2$, $\lambda =1$ case is existence of N\'eel
order at $T=0$ for $\alpha < \alpha^{sw}_c= 2.46$.

In this work, we extend the results of Yusuf {\it et al.} in several
ways, focusing on the zero temperature behavior of the non frustrating
spin $1/2$ Hamiltonian (\ref{QLRO}) with long range interaction of
adjustable strength $\lambda$ and exponent $\alpha$. In Sec. II, we
consider the relevance of the long range term as a
perturbation  to the
nearest neighbor interaction,  using a simple heuristic
argument of mean-field type as well as the power-counting of the
scaling dimension of the perturbation.  For
$\lambda\ll 1$, we find that the long range perturbation is marginal
if $\alpha=2$ and relevant (irrelevant) for $\alpha<2$ ($\alpha>2$).  We then
investigate the  $\alpha$- and $\lambda$-dependence of 
the critical behavior using various techniques. We begin, in Sec. III, with
semi-classical calculations: the SW expansion and a large-N
approximation based on the non-linear $\sigma$ model. Both
approximations give qualitatively similar phase boundaries, 
and sublinear dispersion like in Eq.~(\ref{SL}) in the ordered
phase. Some of the critical exponents can also be estimated within
these approximations.  However, the results obtained in the SW or
large-N approximations are not 
quantitatively correct. We therefore use large scale numerical
simulations to investigate more precisely the phase diagram of this
model in Secs. IV and V. We study  systems of up to
$L=4000$ sites using quantum Monte Carlo (QMC) methods, based on a
stochastic series expansion (SSE) of the partition
function~\cite{Sandvik02,Sandvik03}. We verify that for $S=1/2$, there
are indeed stable phases with both QLRO given by Eq. (\ref{corr}) and
with true N\'eel LRO [Eq.~(\ref{Neel})]. We accurately
determine the  phase boundary, as well as some of the
critical exponents which are found to vary continuously along the
critical line.  In Sec. VI, we also apply analytic renormalization
group (RG) methods to
investigate the case $\lambda \ll 1$.  Sec. VII contains 
conclusions. In two appendices we 
gives further details on the spin-wave theory and large-N calculations. 

\section{Relevance of the perturbation: mean field and scaling
  arguments}
\label{sec:heur}
Let us consider a short range spin $1/2$ chain with an
additional long range perturbation of the form
\be
\label{pert}
\sum_{r,r'}J(r,r'){\vec S}_{r}\cdot {\vec S}_{r'},
\ee
with
\be 
\label{jex}
J(r,r')=-\frac{(-1)^{|r-r'|}}{|r-r'|^{\alpha}}.
\ee
Following an argument given by Cardy~\cite{CardyBook} for the
relevance of a long range perturbation, we can in first approximation
look at the mean field correction to the free energy coming
from this long range term (\ref{pert}):
\be
\delta F =\sum_{r,r'}J(r,r')\langle {\vec S}_{r}\cdot {\vec
  S}_{r'}\rangle,
\ee
where $\langle ... \rangle$ is evaluated in the unperturbed system
where we know the behavior of the correlation function
\be
\langle {\vec S}_{r}\cdot {\vec S}_{r'}\rangle \sim
\frac{(-1)^{|r-r'|}}{|r-r'|^{z+\eta-1}}.
\ee
In a finite system of length $L$, the change
in the free energy per site $\delta f$ thus scales like
\be
\delta f  \sim \int_{1}^{L}\frac{dr}{r^{\alpha+z+\eta-1}}.
\ee
The integral above will give a constant term and a size dependent term
\be
\delta f(L)\sim  L^{2-\alpha-z-\eta}\sim L^{-\alpha},
\ee
where we have used the fact that $z=\eta=1$ in the short range QLRO
regime of the spin $1/2$ chain. Then, we
  can compare this with the usual 
finite size corrections to the free energy of the conformally
invariant short range $S=1/2$ chain which are known to scale like $L^{-2}$ 
to lowest order~\cite{Ian86}.  This tells us that (to first order
perturbation) if $\alpha<2$ the long range perturbation creates a
correction which dominates the $L^{-2}$ correction of the unperturbed
fixed point and is probably a relevant perturbation for the short
range model.

Another way of deriving this result is to compute the scaling
dimension of the perturbation, based on the usual continuum
formulation  of the short range model in which uniform and staggered
magnetization density operators, $(\vec J_L +\vec J_R)$ and $\vec n $
are introduced:
\be
\vec S(x) \approx (\vec J_L+\vec J_R)+(-1)^x\vec
n(x).\label{cont_lim}
\ee
Only slowly varying Fourier modes of the fields $\vec
J_{L/R}(x)$ and $\vec n(x)$ are present in the low energy effective
Hamiltonian. $\vec J_{L/R}$ are the conserved left/right-moving spin
densities. Ignoring a marginally irrelevant interaction, the staggered
magnetization field, $\vec n$, has the Green's function:
\be
\langle n^a(z)n^b(0)\rangle = {\delta^{ab}\over |z|},
\label{ncorr}
\ee
with $z\equiv \tau + ix$.
The long range perturbation adds to the low energy, continuum
limit of the imaginary
time action, a term of the form:
\be
\delta S[\vec n]\sim -\lambda\int {d\tau dx dy\over
|x-y|^{\alpha}}\vec n(\tau ,x)\cdot \vec n(\tau ,y).
\label{Seff}
\ee
Utilizing the fact that, from Eq. (\ref{ncorr}),
$\vec n$ has a scaling dimension of 1/2, a simple power counting tells
us that the perturbation is irrelevant for $\alpha >2$, relevant for
$\alpha <2$, and
marginal for $\alpha =2$.
Also note that $\lambda >0$ corresponds to
non-frustrating interactions which favor the N\'eel state with
$\langle n^z\rangle\neq 0$.
\section{Spin-wave expansion and Large-N approximation}
\label{sec:SW}
\subsection{Spin-wave expansion}
This calculation simply generalizes  that of Yusuf {\em et al.} in
Ref. \cite{Yusuf04}, to $\lambda \ne 1$. We summarize here the main
steps. Some further results are  given in Appendix A. On the LRO side of
the transition, we use the 
Holstein-Primakoff approximation~\cite{Holstein}: 
$$
S_{i}^z = S-a^{\dagger}_{i}a_{i}, S_{i}^{+}\approx
\sqrt{2S} a_{i}, S_{i}^{-}\approx
\sqrt{2S} a^{\dagger}_{i}, 
$$ 
for $i$ odd and 
$$
S_{j}^z = b^{\dagger}_{j}b_{j}-S, S_{j}^{+}\approx
\sqrt{2S} b^{\dagger}_{j}, S_{j}^{-}\approx
\sqrt{2S} b_{j}, 
$$ 
for $j$ even, and  retain only the quadratic terms in the Hamiltonian (\ref{QLRO}). 
After a Fourier transform over the reduced Brillouin
zone $k\in (-{\pi\over 2a}, {\pi\over 2a})$, we find:
\begin{equation}
{\cal{H}}_{SW} \approx S \sum_{k}^{}\Bigl[ \left(\gamma - f(k)
  \right)\left(a_k^{\dagger} a_k + b_k^{\dagger} b_k  \right)
+ g(k) \left(a_k^{\dagger} b_{-k}^{\dagger} +b_{-k} a_k\right)\Bigr]+\dots
\ee
where, for an infinite chain \cite{note2}:
$$
\gamma = 2 + 2 \lambda \sum_{n=2}^{\infty} \frac{1}{(2n-1)^{\alpha}}
$$
$$
f(k) = 2  \lambda \sum_{n=1}^{\infty} \frac{ \cos(2kna)-1}{(2n)^{\alpha}}
$$
$$
g(k) = 2 \cos(ka) + 2  \lambda \sum_{n=2}^{\infty}
\frac{\cos[k(2n-1)a]}{(2n-1)^\alpha} 
$$
This quadratic Hamiltonian can be diagonalized with a 
Bogoliubov transformation to:
\be
{\cal H}_{SW} \approx S \sum_{k}^{} \omega_k
\left(\chi_{k,1}^{\dagger}\chi_{k,1}+\chi_{k,2}^{\dagger}\chi_{k,2}
\right) 
\label{hsw}
\ee
with a SW spectrum 
\be
 \omega_k = \sqrt{\left[\gamma - f(k)
  \right]^2 + [g(k)]^2 }\stackrel{k\rightarrow 0}{\longrightarrow}
k^{\alpha-1\over 2} , 
\ee
as discussed above. At $T=0$, the correction
to the staggered magnetization at any site is
$$
\Delta m_q = \langle a^{\dagger}_{i} a_{i}\rangle = \langle
b^{\dagger}_{j} b_{j}\rangle = {a\over 2\pi} \int_{-{\pi\over
	2a}}^{\pi\over 2a} dk \left[\frac{\gamma - f(k)}{\omega_k} -1\right]
$$
The consistency condition  $\Delta m_q
< S$ then allows us to find the SW approximation for the value
of $\alpha_{c}^{sw}$ below which long range N\'eel order is established. As
already stated, for  $S=1/2$ and $\lambda =1$, we find $\alpha_{c}^{sw} =
2.46$. A plot of $\alpha_{c}^{sw}$ vs. $\lambda$, for $S=1/2$, is shown in
Fig.~\ref{fig:SW}. 

\begin{figure}[!ht]
\bc
\includegraphics[width=8cm,clip]{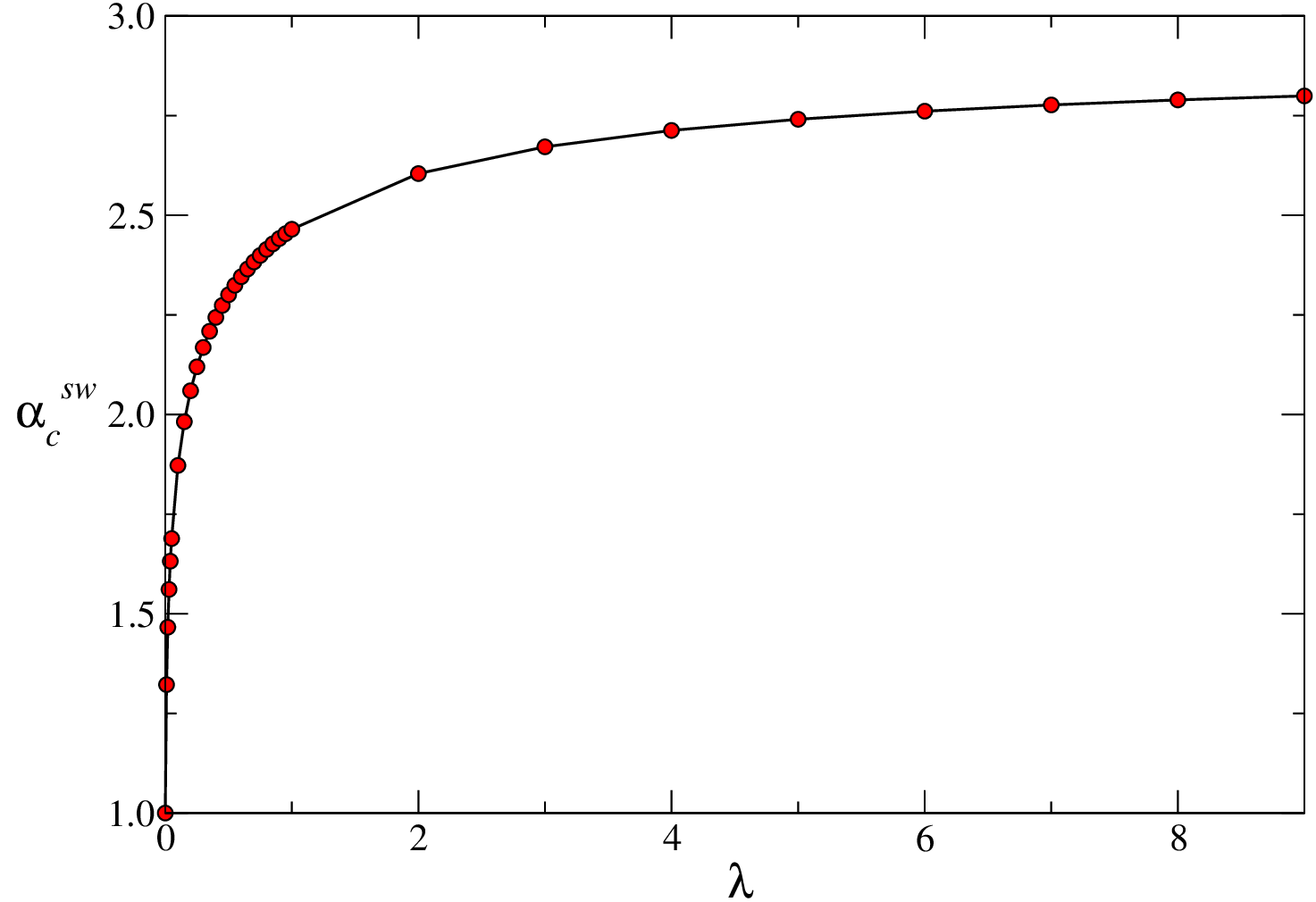}
\caption{Spin-wave approximation prediction for the value
  $\alpha_{c}^{sw}$ below which long range N\'eel order is expected at
  $T=0$, as a function of $\lambda$ and for $S=1/2$. The critical
  curve saturates at $\alpha \simeq 2.9032$ when $\lambda\to\infty$.}
\label{fig:SW}
\ec
\end{figure}
This phase boundary given by the lowest order SW approximation turns
out not to be quantitatively correct, as we are going to see with the QMC
results presented in section~\ref{sec:QMC1}. In particular it happens that SW
predictions miss the fact that the critical line goes to $\alpha=2$
when $\lambda \to 0$. Nevertheless SW predictions are, to some
extent, shared by large-$N$ calculations as we are going to see
below.
    
\subsection{Large-$N$ approximation}
The details of this calculation are given in
Appendix~\ref{app:LN}. Here we present the main steps and discuss the
results which come from this approximation. We generalize the 
N\'eel order parameter field, $\vec n(\tau,x)$ appearing in Eq. (\ref{cont_lim}), 
to an $N$-component field and take the limit of large $N$. In this 
approximation 2 phases occur in the $\lambda$ -$\alpha$ plane. 
The critical line terminates at $\alpha =1$, as in spin-wave approximation. 
These two phases  are 
a phase with N\'eel order and a disordered phase with a finite correlation length. 
(The unusual quasi long range ordered phase is special to the case $N=3$ 
and is not captured by the large-$N$ approximation.) Along the critical line 
separating these two phases the mean field result, $\eta =3-\alpha$ is obtained. 
The dynamical exponent takes the value $z=(\alpha -1)/2$, corresponding 
to the dispersion relation $\omega \propto |k|^{(\alpha -1)/2}$ also 
obtained in spin-wave theory. The correlation length diverges with an exponent, $\nu$ 
defined by:
\be \xi \propto |\lambda_c-\lambda|^{-\nu},\ee
with 
\bea \nu &=& 1/(\alpha -1),\ \ (1<\alpha <5/3) \nonumber \\
\nu &=& 2/(3-\alpha ),\ \  (5/3<\alpha <3).\eea

%%%%%%%%%%%%%%%%%%%%%%%%%%%%%%%%%%%%%%%%%%%%%%%%%%%%%%%%
\section{Quantum Monte Carlo results I: Finite size effects}
%%%%%%%%%%%%%%%%%%%%%%%%%%%%%%%%%%%%%%%%%%%%%%%%%%%%%%%%
\label{sec:QMC1}
In this section, we present results obtained using the QMC SSE method
based on directed loop
updates~\cite{Sandvik02}. This algorithm, used here to investigate the
model (\ref{QLRO}), has been proposed recently by Sandvik~\cite{Sandvik03}
to study spin Hamiltonians with non-frustrating long range interactions.
%%%%%%%%%%%%%%%%%%%%%%%%%%%%%%%%%%%%%%%%
\subsection{Finite size corrections}
%%%%%%%%%%%%%%%%%%%%%%%%%%%%%%%%%%%%%%%%

We first focus on the $\lambda=1$ case, studied by SW in Ref.~\cite{Yusuf04}, governed by the following
Hamiltonian
\be
{\cal H}=-\sum_{i,j\geq 1}\frac{(-1)^j}{j^{\alpha}}{\vec S}_{i}\cdot{\vec S}_{i+j}.
\label{eq:iso}
\ee
In order to detect a N\'eel instability at the thermodynamic limit, 
we compute the staggered structure factor, normalized per site, on finite length
spin $S=1/2$ chains, defined by
\be
S_{\pi}(L)=\frac{1}{L^2}\sum_{i,j}(-1)^{i-j}\langle{{\vec S}}_{i}\cdot{{\vec S}}_{j}\rangle=\frac{3}{L^2}\langle (\sum_{i=1}^{L}(-1)^i S_{i}^{z})^2\rangle.
\ee
We have performed SSE simulations for different system sizes, up to
$L_{\rm{max}}=4096$, at temperatures $\beta^{-1}=1/2L$
low enough to get the GS properties.
Results for $S_{\pi}(L)$ are shown  vs $1/L$ in the left panel of
Fig.~\ref{fig:spi} for
different values of the power-law exponent $\alpha$. The staggered
structure factor displays two types of behavior: for small values of
$\alpha$ it saturates to a finite non-zero number whereas for large
enough $\alpha$, $S_{\pi}(L)$ vanishes when $L\to\infty$.
%%%%%%%%%%%%%%%%%%%%%%%%%%%%%%%%%%%%%%%%
\begin{figure}[!ht]
\bc
\psfrag{B}{\Huge{N\'eel}}
\psfrag{K}{\Large QLRO}
\begin{minipage}{\columnwidth}
\includegraphics[width=0.46\columnwidth,clip]{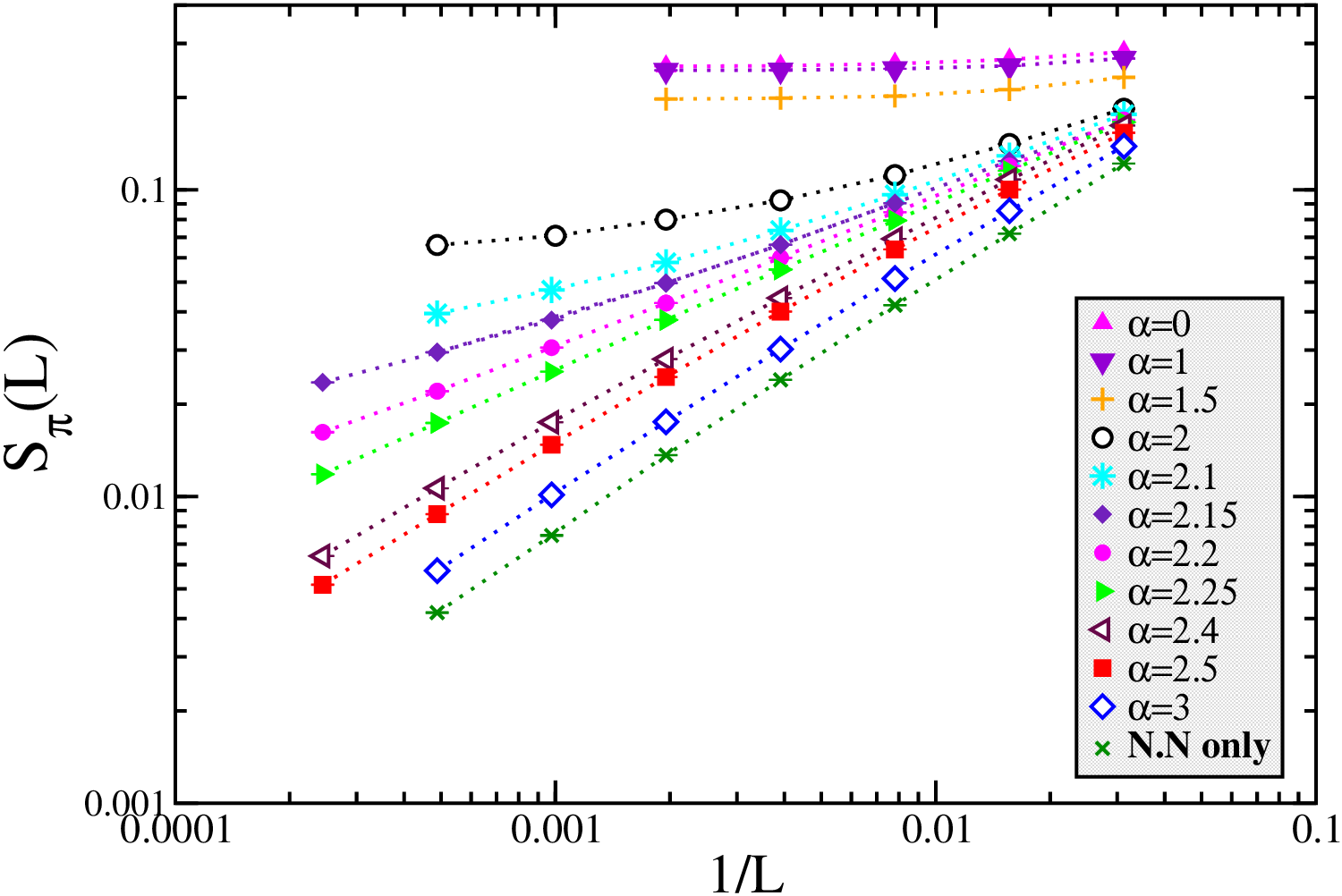}
\includegraphics[width=0.44\columnwidth,clip]{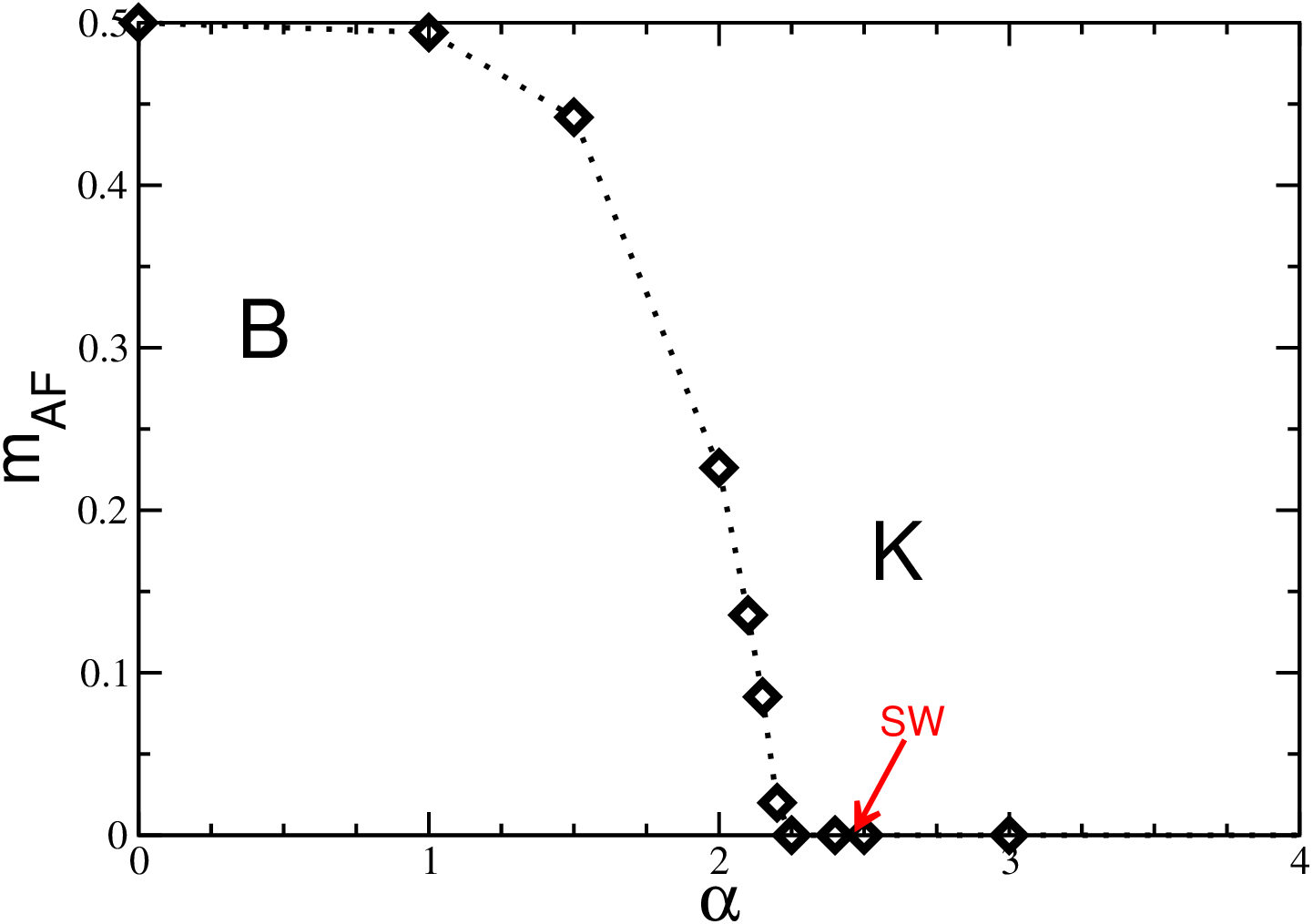}
\end{minipage}
\caption{Left panel: $T=0$ QMC results for the staggered structure factor per
  site $S_{\pi}(L)$ computed in the GS for the Hamiltonian
  (\ref{eq:iso}) and plotted vs the inverse system size
  $L^{-1}$ in a log-log scale.
  Different symbols are used for different values of the
  power-law exponent $\alpha$, as indicated on the plot. The case
  with only nearest neighbor interactions is also shown (N.N green crosses) for
  comparison. Right panel: Infinite size AF order
  parameter $m_{\rm{AF}}$ plotted vs $\alpha$, obtained using finite size scaling of
  $S_{\pi}(L)$ (shown on the left panel). The quantum
  phase transition between the N\'eel
  phase ($m_{\rm{AF}}\ne 0$) and the QLRO phase ($m_{\rm{AF}}=0$)
  occurs at $\alpha_c=2.225\pm 0.025$. The SW estimate
  ($\alpha_{c}^{\rm{sw}}\simeq 2.46$) is indicated by the arrow.}
\label{fig:spi}
\ec
\end{figure}
Then, in order to extract the thermodynamic limit behavior of
$S_{\pi}$, we perform a finite size analysis in order to get the AF
order parameter, given by
\be
\sqrt{S_{\pi}(L)}\to m_{\rm{AF}},\  \ (L\to \infty).
\label{maf1}
\ee
Utilizing the fact that in the QLRO regime the spin-spin correlation
functions decay as stated in Eq.~(\ref{corr}), we therefore expect in
this regime the following behavior for the staggered structure factor
per site
\be
S_{\pi}(L)= \frac{1}{L}\int_{1}^{L}(-1)^r\langle {\vec{S}}_0 \cdot
{\vec{S}}_r \rangle dr\sim \frac{\left(\ln L\right)^{3/2}}{L},\  \ (L\to \infty).
\ee
On the other hand, in the N\'eel phase, the finite size scaling of the
order parameter can be evaluated using the small $k$ SW spectrum (see appendix~\ref{app:SW}), leading to  
\be
S_{\pi}(L)-m_{\rm{AF}}^2\sim L^{\frac{\alpha-3}{2}}+O(L^{{\alpha-3}}).
\label{eq:fss}
\ee 
We used second order polynomial fits in $L^{\frac{\alpha-3}{2}}$ to
  extrapolate the finite size data to their thermodynamic limit
  values, shown in the right panel of Fig.~\ref{fig:spi}.
The quantum phase transition between the AF N\'eel order and the
QLRO phase is clearly visible for a critical value
$2.2<\alpha_c<2.25$. It is also interesting to compare this estimate
with the one from SW approximation giving $\alpha_{c}^{\rm{sw}}\simeq 2.46$.

Let us now concentrate on the $\alpha-$ and $\lambda-$dependent
Hamiltonian (\ref{QLRO}) by keeping a fixed value for  $\alpha$ while
varying $\lambda$. We first focus
on the case with $\alpha=2.1$ which is expected to display a
transition for a non-zero value of $\lambda$.
As pointed out by Reger and Young studying finite size AF clusters in
$d=2$~\cite{Reger88}, the sublattice (infinite size) magnetization can be obtained
either from the staggered structure factor [Eq.~(\ref{maf1})] or from
the correlation functions at the largest separation
\be
\label{maf2}
C(L)=\langle \vec{S}_i\cdot\vec{S}_{i+L/2}\rangle \to
\pm m_{\rm{AF}}^{2},\  \ (L\to \infty).
\ee
In the N\'eel phase, both estimators $S_{\pi}(L)$ and $C(L)$ are
expected to converge to $m_{\rm{AF}}^2$ with a similar 
power-law behavior but with different pre-factors. This feature is
illustrated by the computation of $C(L)$ and $S_{\pi}(L)$ for
$\alpha=2.1$ and four different value of the long range term strength
$\lambda=3,~2~,0.9,~0.6$, as shown in Fig.~\ref{fig:2.1}.
%%%%%%%%%%%%%%%%%%%%%%%%%%%%%%%%%%
\begin{figure}[!ht]
\bc
\psfrag{Z}{\tiny{\bf{(a) $\lambda=3$}}}
\psfrag{Y}{\tiny{\bf{(b) $\lambda=2$}}}
\psfrag{X}{\tiny{\bf{(c) $\lambda=0.9$}}}
\psfrag{W}{\tiny{\bf{(d) $\lambda=0.6$}}}
\includegraphics[width=10cm,clip]{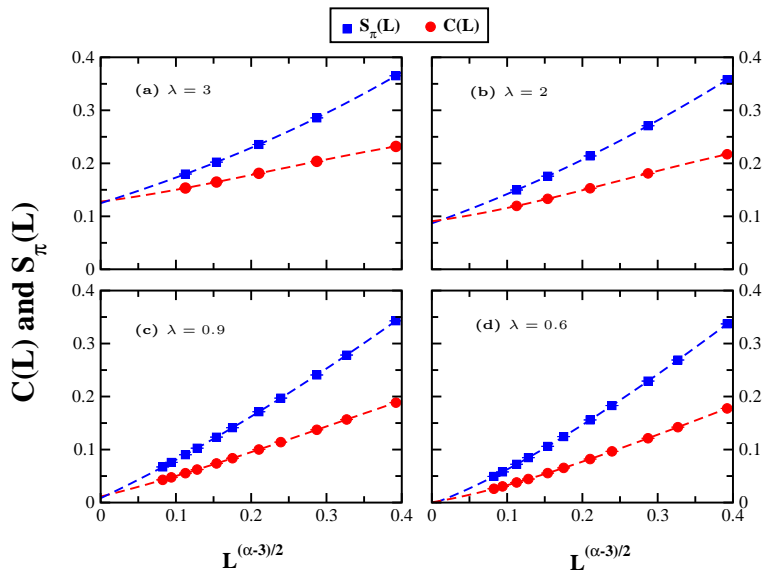}
\caption{Staggered structure factor per site
  $S_{\pi}(L)$ [Eq.~(\ref{maf1})] and mid-chain correlation function
  $C(L)$ [Eq.~(\ref{maf2})] computed at $T=0$ with QMC for $\alpha=2.1$
  and 4 different
  values of $\lambda$, as shown on the plot (a)-(d). Finite size
  scaling has been performed on finite systems up to
  $L_{\rm{max}}=128$ for the largest values of $\lambda$ ((a) and
  (b)) and up to $L_{\rm{max}}=256$ for the smallest ((c) and
  (d)). The dashed lines are polynomial fits of the form
  $m_{\rm{AF}}^2+a_1 L^{\frac{\alpha-3}{2}}+a_2 L^{\alpha-3}$.}

\label{fig:2.1}
\ec
\end{figure}
%%%%%%%%%%%%%%%%%%%%%%%%%%%%%%%%%
Using second order polynomial fits in $L^{\frac{\alpha-3}{2}}$, we can
obtain infinite size extrapolated values for $m_{\rm{AF}}(\lambda)$
from $S_{\pi}(L)$ or $C(L)$, as reported in table~\ref{table1}.
Finite size effects are more pronounced for the staggered structure
factor than for the mid-chain correlation function because
$S_{\pi}(L)$ is the result of the integration of the staggered
correlation function along the entire chain and therefore is sensitive
to short distance terms.  However, the estimates for the sublattice
magnetization obtained from $C(L)$ and $S_{\pi}(L)$ (see
table~\ref{table1}) are both in good agreement, especially when the
system is deeply in the N\'eel regime (large values of $\lambda$). On
the other hand, when the system is approaching the quantum critical
point (QCP) where $m_{\rm{AF}}\to 0$, the finite size effects are
significant enough to prevent us from obtaining a very precise
estimate of the critical coupling $\lambda_c$ where the AF LRO
vanishes.

\begin{table}
\label{table1}
\begin{tabular}{r|r|r|}
$\lambda$&$m_{\rm{AF}}$ from $S_{\pi}(L)$&$m_{\rm{AF}}$ from $C(L)$\\
\hline
3&0.353&0.356\\
2&0.295&0.301\\
0.9&0.106&0.091\\
0.6&0&0.02\\
\end{tabular}
\caption{Infinite size extrapolated values of the
sublattice magnetization $m_{\rm{AF}}$ obtained for $\alpha=2.1$ and
$\lambda=3,~2,~0.9,~0.6$ from power-law fits of the staggered
structure factor $S_{\pi}(L)$ and the mid-chain correlation function $C(L)$ (see Fig.~\ref{fig:2.1}).}
\end{table}
%
%\begin{table}
%\begin{indented}
%\item[]\begin{tabular}{@{}lll}
%\br
%$\lambda$&$m_{\rm{AF}}$ from $S_{\pi}(L)$&$m_{\rm{AF}}$ from $C(L)$\\
%\mr
%3&0.353&0.356\\
%2&0.295&0.301\\
%0.9&0.106&0.091\\
%0.6&0&0.02\\
%\br
%\end{tabular}
%\caption{\label{table1}Infinite size extrapolated values of the
%sublattice magnetization $m_{\rm{AF}}$ obtained for $\alpha=2.1$ and
%$\lambda=3,~2,~0.9,~0.6$ from power-law fits of the staggered
%structure factor $S_{\pi}(L)$ and the mid-chain correlation function $C(L)$ (see Fig.~\ref{fig:2.1}).}
%\end{indented}
%\end{table}
%
Of course, in principle it is  possible to perform
very large scale numerical SSE simulations on the largest reachable system
sizes, as we did for the $\lambda=1$ case with
$L_{\rm{max}}=4096$. However, since our goal here is to investigate
the quantum critical phenomena in the $\lambda-\alpha$ plane, we need
a good sampling of this parameter space and we therefore restrict
the simulations over systems of maximum size $L_{\rm{max}}\le 1024$.
We then  use another strategy, based on scaling
arguments, to perform a better data analysis
close to criticality. This is described next.
%%%%%%%%%%%%%%%%%%%%%%%%%%%%%%%%%%%%%%%%
\subsection{Scaling analysis}
%%%%%%%%%%%%%%%%%%%%%%%%%%%%%%%%%%%%%%%%
As previously discussed, the finite size effects are bigger for the
staggered structure factor than for the mid-chain correlation
function. Therefore, we now focus on $C(L)$ which is
expected to saturate to a constant value in the N\'eel phase, whereas in the
QLRO regime, the  behavior $C(L)\to \sqrt{\ln (L/a)}/L$ is
expected, $a$ being a non-universal constant.

%%%%%%%%%%%%%%%%%%%%%%%%%%%%%%%%%%%%%%%%
\begin{figure}[!ht]
\bc
\psfrag{W}{$C(L)$}
\psfrag{X}{$L$}
\psfrag{Y}{$\Psi_{\lambda}C(L)$}
\psfrag{Z}{$L/\xi_{\lambda}$}
\begin{minipage}{\columnwidth}
{\hskip -0.2cm{\includegraphics[width=0.49\columnwidth,clip]{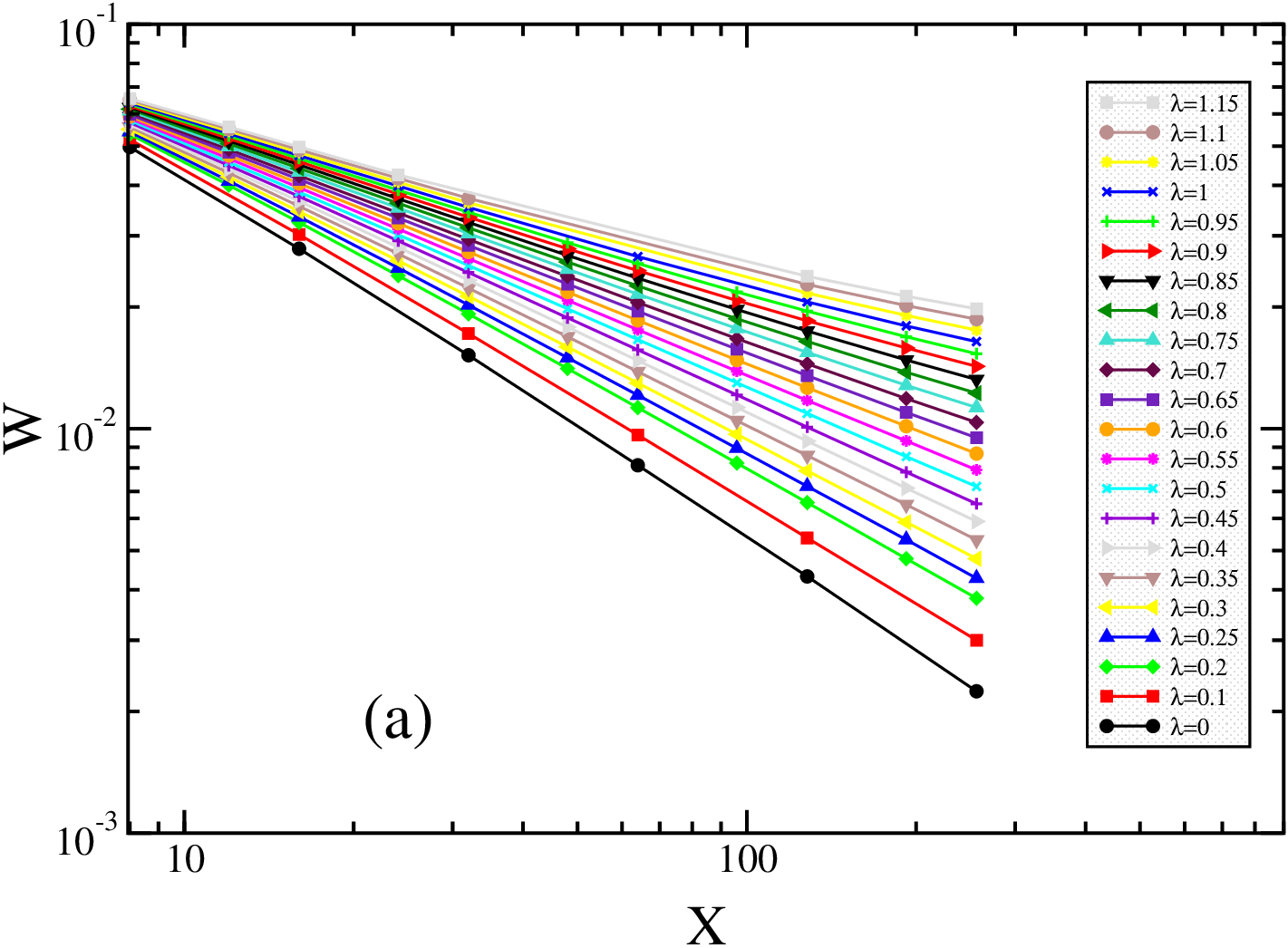}}}
{\hskip 0.2cm{\includegraphics[width=0.49\columnwidth,clip]{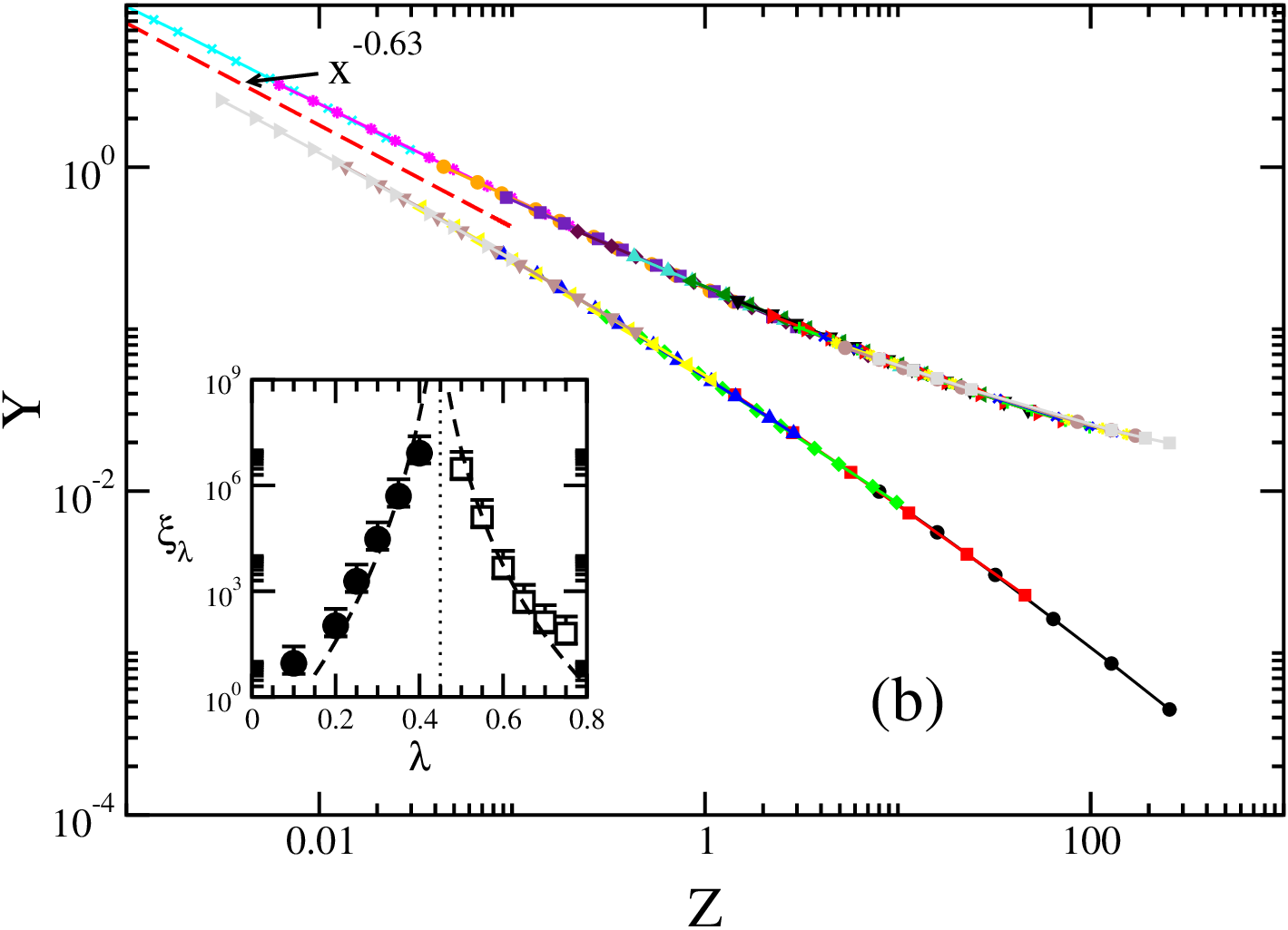}}}
\end{minipage}
\caption{$T=0$ QMC results for the mid-chain correlation function $C(L)$
  computed for $\alpha=2.1$ and different values of $\lambda$, as
  indicated on the plot. (a) $C(L)$ is plotted vs the system size $L$ for
  $0\le \lambda\le 1.15$. (b)
  Both x- and y-axis are rescaled using two parameters:
  the crossover
  length scale $\xi_{\lambda}$ and $\Psi_{\lambda}$. The data collapse results in two
  universal curves: one for the N\'eel ordered phase (top one) and
  one for the QLRO regime (lower one).  Note that for clarity, the
  QLRO universal curve has been shifted downwards. The red dashed-line
  materializes the critical separatrix between the two regimes, decaying with
  an exponent $\simeq 0.63$. Inset: Crossover
  length scale $\xi_{\lambda}$ extracted from the data collapse in the
  QLRO (full circles) and N\'eel regime (open squares). The dashed lines
  are power-law fits of the form $|\lambda-\lambda_c|^{-\nu}$
  with $\lambda_c\simeq 0.45$ (indicated by vertical
  dotted line) and $\nu\sim 15$.}
\label{fig:cL}
\ec
\end{figure}
%%%%%%%%%%%%%%%%%%%%%%%%%%%%%%%%%
In order to illustrate the scaling analysis, let us continue to study
the case with $\alpha=2.1$, as in the previous 
subsection. We have computed $C(L)$ for several values of the long range coupling
strength $\lambda$ in the range $[0,1.15]$, for sizes up to $L=256$. The results,
shown in Fig.~\ref{fig:cL} (a), clearly show the existence of a
finite critical value
$\lambda_c$ which separates the QLRO and the N\'eel regimes. In order to
locate precisely this QCP, let us assume that
a typical length 
scale $\xi_{\lambda}$ governs a crossover from the QCP to the N\'eel
phase if $\lambda > \lambda_c$ and from the QCP to the QLRO regime if
$\lambda < \lambda_c$. Precisely at the critical point, the spin-spin
correlation function decay like a power-law
\be
C_{\rm{QCP}}(L)\sim L^{1-z-\eta},
\label{eq:QCP}
\ee
thus defining the critical exponents $\eta$ and $z$, the critical
dynamical exponent. Without making any 
assumption about the values of the aforementioned critical exponents,
let us now define  scaling 
functions $f_{\pm}(x)$, with $x=L/\xi_{\lambda}$,
for $\lambda>\lambda_c$ and $\lambda<\lambda_c$, respectively, by
\be
f_{\pm}(x)=\frac{C(L)}{C_{\rm{QCP}}(L)}
\ee
Hence, the scaling functions obey
\bea
\label{eq:scaling}
f_-(x)&\propto&
x^{-2+z+\eta}{\sqrt{\ln x}}{\rm{~if~}} x\gg 1 {\rm{~and~}}   \lambda
< \lambda_c {\rm{~~~~(QLRO)}}\nonumber \\ 
f_+(x)&\propto&x^{-1+z+\eta}{\rm{~if~}} x\gg 1 {\rm~{and~}} \lambda > \lambda_c {\rm{~~~~(NEEL)}}\nonumber \\
f_{\pm}(0)&=&1  {\rm{~if~}} \lambda \simeq \lambda_c {\rm{~~~~(QCP)}}
\eea
It is convenient to also rescale the y-axis with the unknown function
$\Psi_{\lambda}$ in order to get 
$C(L)\times
\Psi_{\lambda}=f(x)\times x^{1-z-\eta}$.
We then expect $\Psi_{\lambda}$ to be proportional to $\xi_{\lambda}^{z+\eta-1}$.
Using such scaling forms, we have obtained the collapse of the data
shown in Fig.~\ref{fig:cL}(a) into two universal curves shown in
Fig.~\ref{fig:cL}(b). The parameters $\xi_{\lambda}$ and
$\Psi_{\lambda}$ have been chosen to give the best data collapses.
Using such a scaling analysis, we find a critical coupling
$\lambda_c=0.45\pm 0.05$ that we can compare to the overestimated value
$\lambda_c=0.6$ previously found using the more simple finite size
scaling Eq.~(\ref{eq:fss}). The critical correlation (given by the separatrix between the two
regimes in Fig.~\ref{fig:cL}(b)) is characterized here by a power-law
decay with an exponent $(z+\eta-1)_{\rm{QCP}}\simeq 0.63$.
Note also that the crossover length scale $\xi_{\lambda}$, plotted in the inset of
Fig.~\ref{fig:cL}(b), diverges on both sides of the
transition with
a large exponent $\nu\sim 15$~\cite{notexi}.
These and other issues related to the critical exponents will be discussed in
detail in Sec.~\ref{sec:crit}.
%%%%%%%%%%%%%%%%%%%%%%%%%%%%%%%%%%%%%%%%
\section{Quantum Monte Carlo II: Phase diagram and Critical behavior}
%%%%%%%%%%%%%%%%%%%%%%%%%%%%%%%%%%%%%%%%
The scaling analysis described above has been repeated for several values
of $\alpha$ in order to explore and construct the phase diagram of the
model (\ref{QLRO}) in the $\lambda-\alpha$ plane. 
%%%%%%%%%%%%%%%%%%%%%%%%%%%%%%%%%%%%%%%%
\subsection{$\alpha=2$: Marginal case}
%%%%%%%%%%%%%%%%%%%%%%%%%%%%%%%%%%%%%%%%
Let us first focus on the marginal case with
$\alpha=2$ for which a similar data collapse analysis is performed and
shown in Fig.~\ref{fig:cL1} for the mid-chain correlation
function. 

%%%%%%%%%%%%%%%%%%%%%%%%%%%%%%%%%%%%%%%%
\begin{figure}[!ht]
\bc
\psfrag{Y}{$\Psi_{\lambda}C(L)$}
\psfrag{Z}{$L/\xi_{\lambda}$}
\includegraphics[width=8cm,clip]{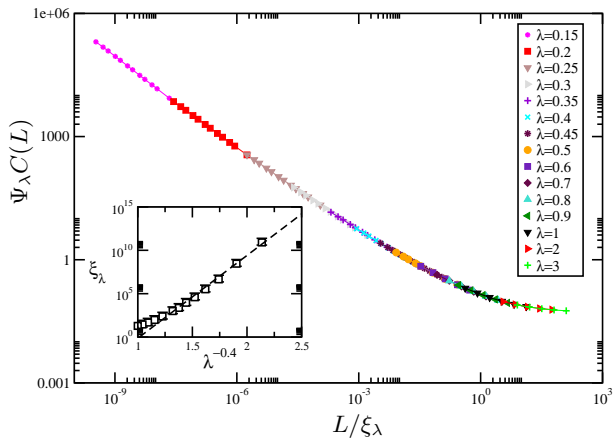}
\ec
\caption{$T=0$ QMC results at $\alpha=2$ for the mid-chain correlation
  functions $C(L)$. As in
  Fig.~\ref{fig:cL}, both x- and y-axis have been rescaled in order
  to get the best data collapse. For the different values of $\lambda$
  indicated on the plot, the data collapse on a unique crossover
  curve towards the N\'eel ordered phase. Inset: Crossover
  length scale $\xi_{\lambda}$ plotted in a linear-log scale vs
  $\lambda^{-0.4}$. The dashed line is a fit of the form
  Eq.~(\ref{eq:expfit}) with $\sigma=0.4$.}
\label{fig:cL1}
\end{figure}
%%%%%%%%%%%%%%%%%%%%%%%%%%%%%%%%%%%%%%%%
Using QMC simulations results for chains up to $L=512$ sites, with
$\lambda \in [0,3]$, we have been able to get an universal curve (see
Fig.~\ref{fig:cL1}) which shows a crossover towards a N\'eel order
phase (i.e. $C(L)\to{\rm{constant}}$ if $L\gg \xi_{\lambda}$). 
Note that for $\lambda < 0.1$, the typical
length scale necessary to get a good collapse becomes very large 
so that there is no overlap between our different curves and the 
data collapse analysis is impossible to achieve.
Nevertheless, the crossover length scale
$\xi_{\lambda}$, plotted in the inset of Fig.~\ref{fig:cL1}, displays
an exponential divergence when $\lambda\to 0$. Guided by the RG calculations presented in
section~\ref{sec:RG}, we can fit the $\lambda$-dependence of the crossover length scale by
\be\label{eq:expfit}
\xi_{\lambda}\sim \exp(C/\lambda^{\sigma}),
\ee
with $\sigma=0.4$ and $C$ being a free parameter.
It is however important to note
that since $\xi_{\lambda}$ suffers from large error bars, and so does the
fitting parameter, we have
forced $\sigma$ to its value found in Eq.~(\ref{eq:exp}). 

Unambiguously, $\xi_{\lambda}$ is found to diverge when $\lambda\to 0$
which means that at the marginal point
$\alpha=2$, any $\lambda >0$ will drive the system towards the N\'eel
phase. In other words, the
long range interaction perturbation of strength $\lambda$ is
{\it{marginally relevant}} at $\alpha=2$. This result agrees 
with the RG calculations presented
in section~\ref{sec:RG}.
%%%%%%%%%%%%%%%%%%%%%%%%%%%%%%%%%%%%%%%%
\subsection{Phase Diagram}
%%%%%%%%%%%%%%%%%%%%%%%%%%%%%%%%%%%%%%%%
As previously stated, when $\alpha\le 2$ the long range interaction is a
relevant perturbation and any $\lambda>0$ will drive the QLRO
phase towards a AF ordered N\'eel phase with
$m_{\rm{AF}}\ne 0$. On the other hand, when $\alpha> 2$ a simple
power-counting tells us that the perturbation is irrelevant which
should imply that the QLRO is stable against a small perturbation
$\lambda>0$. 
%%%%%%%%%%%%%%%%%%%%%%%%%%%%%%%%%%%%%%%%
\begin{figure}[!ht]
\bc
\psfrag{NEEL}{\Huge N\'eel}
\psfrag{QLRO}{\Huge QLRO}
\includegraphics[width=10cm,clip]{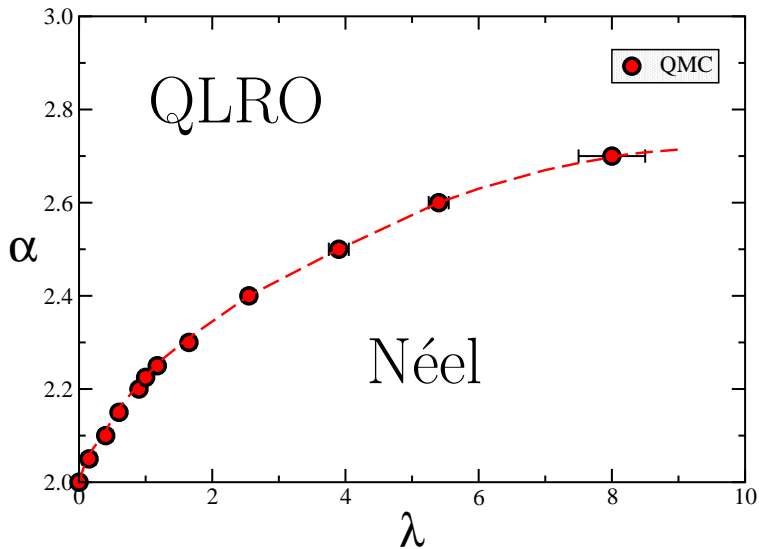}
\caption{$T=0$ phase diagram of the long range $S=1/2$ model
  [Eq.~(\ref{QLRO})] computed by large scale QMC
  simulations and plotted in the $\lambda-\alpha$ plane. A line of
  critical points (circles) separates a long range ordered phase
  (N\'eel) and a quasi long range ordered phase (QLRO). The error
  bars, due to some uncertainties in the finite size scaling analysis of
  the numerical data, are explicitly show on the plot. The dashed line
  is a guide for the eyes.}
\label{fig:phasediag}
\ec
\end{figure}
%%%%%%%%%%%%%%%%%%%%%%%%%%%%%%%%%%%%%%%%%
It turns out that such a simple argument is not sufficient to provide
a correct description of the quantum critical behavior of the
system (see the next section for a more advanced field theoretical
description). Based on large scale numerical simulations, we
provide  hereafter a picture which is consistent with the existence of a
non-trivial line of fixed points in the 
$\lambda-\alpha$ plane.

Using QMC simulations on systems of up to $L=1000$, we 
performed the scaling analysis for the mid-chain correlation
function $C(L)$ as well as for the staggered susceptibility (see
below) and computed the phase diagram for $2\le \alpha\le
2.7$. For each value of $\alpha$, the QCP $\lambda_c$ is found
by the separatrix between the two crossover functions (see
Fig.~\ref{fig:cL}(b)) with some error bars due to the discrete
sampling in the $\lambda$ space as well as the strong divergence of
the crossover length scale close to the critical point which makes
the data collapse delicate.
We present in Fig.~\ref{fig:phasediag} the QMC phase diagram in the
$\lambda-\alpha$ plane. 
As discussed, $\lambda$ is marginally relevant at $\alpha =2$, driving 
the system towards a N\'eel phase with LRO. At small $\lambda$, the critical line
increases sharply from $\alpha=2$ and displays a
negative curvature. By contrast, spin-wave theory (see
Fig.~\ref{fig:SW}) and large-$N$
 approximation predict 
that $\alpha_c(\lambda\rightarrow 0 )=1$. In the range of $\lambda$ considered here
($\lambda<8$), the critical line stays well below the value
$\alpha=3$ and  we expect this feature to remain true for all
$\lambda$. This behavior  is
consistent with the proof of absence of LRO at $T=0$ for $\alpha >3$
\cite{Parreira97,note1}.

%
%Let us give a few words about this computed phase boundary in the
%$\lambda-\alpha$ plane which is quantitatively different from the one
%computed with SW (see Fig.~\ref{fig:SW}). 
%First of all, $\lambda$ is marginally relevant at $\alpha =2$, driving 
%the system towards a N\'eel phase with LRO. At small $\lambda$, the critical line
%increases sharply from $\alpha=2$ and displays a
%negative curvature. By contrast, spin-wave theory and large-$N$
% approximation predict 
%that $\alpha_c(\lambda )$ hits the $\alpha$ axis 
%at $\alpha =1$. In the range of $\lambda$ considered here
%($\lambda<8$), the critical line stays well below the upper bound
%$\alpha=3$ and regarding the shape of
%the curve, we expect this feature to remain true.
%Also, we
%could perhaps expect the critical line to eventually saturate asymptotically
%towards a value $\alpha^{*}<3$. This issue is of course only
%speculative and not conclusive at all.

%%%%%%%%%%%%%%%%%%%%%%%%%%%%%%%%%%%%%%%%
\subsection{Critical exponents}
%%%%%%%%%%%%%%%%%%%%%%%%%%%%%%%%%%%%%%%%
\label{sec:crit}
The transition line between N\'eel LRO and QLRO is a
non trivial line which displays continuously varying critical
exponents, as we show now.

\subsubsection{Divergence of the crossover length scale}
\label{sec:nu}
The standard theory of quantum phase transitions involves a set of
critical exponents which govern the universal behavior of various
quantities close to or at the QCP. One of them is 
$\nu$ which tells us how does the correlation length diverge in the
real space direction close to the critical point. Usually this
correlation length is defined in the disordered phase by the
exponential decay $\sim
\exp(-r/\xi)$ of the correlation function associated with the order
parameter. In our case,  the non ordered regime $\lambda<\lambda_c$ is already
critical and thus the correlation length is intrinsically
infinite. Nevertheless, the typical length scale $\xi_\lambda$ which
governs the crossover phenomenon, diverges at the QCP (on both sides) with an exponent which we call $\nu$ by analogy:
\be
\xi_{\lambda} \propto |\lambda -\lambda_c|^{-\nu}.
\ee
As already discussed, an accurate numerical evaluation of the
exponent $\nu$ is difficult, because of some intrinsic
uncertainties in the data collapse procedure.
Nevertheless, at our
level of precision we observe this crossover
length scale exponent increasing when $\alpha\to 2^+$. 
In particular,
at the marginal point $\alpha=2$, we find an exponential divergence
of $\xi_{\lambda}$ near $\lambda=0$ [Eq.~(\ref{eq:expfit})], formally
corresponding to $\nu=\infty$. This divergence of $\nu$ when
$\alpha$ approaches 2 is actually in good agreement with the results of
 field theory and RG calculations presented in Sec.~\ref{sec:RG}.

\subsubsection{Staggered magnetization exponent and hyperscaling relation}
The other scaling parameter $\Psi_{\lambda}$,  used for the collapse of
the correlation functions data, also contains some
information. First, if the scaling hypothesis used above with the help
of the crossover functions is correct, we expect $\Psi_{\lambda}\sim
\xi_{\lambda}^{z+\eta-1}$ which gives another estimate for the critical
exponent of the decay of the
correlation function [Eq.~({\ref{eq:QCP})].  This is illustrated in Fig.~\ref{fig:PsiXi}
where $\Psi_{\lambda}$ is plotted vs $\xi_{\lambda}$ for
$\alpha=2.1$. Data, presented for both sides of the transition in
Fig.~\ref{fig:PsiXi}, clearly display power-law dependences with an
exponent $\eta+z-1$ in very good agreement with the value of $0.63$
previously found along the separatrix in Fig.~\ref{fig:cL}(b). Note
also that the agreement is even better when getting closer to the QCP.
\begin{figure}[!ht]
\bc
\includegraphics[width=8cm,clip]{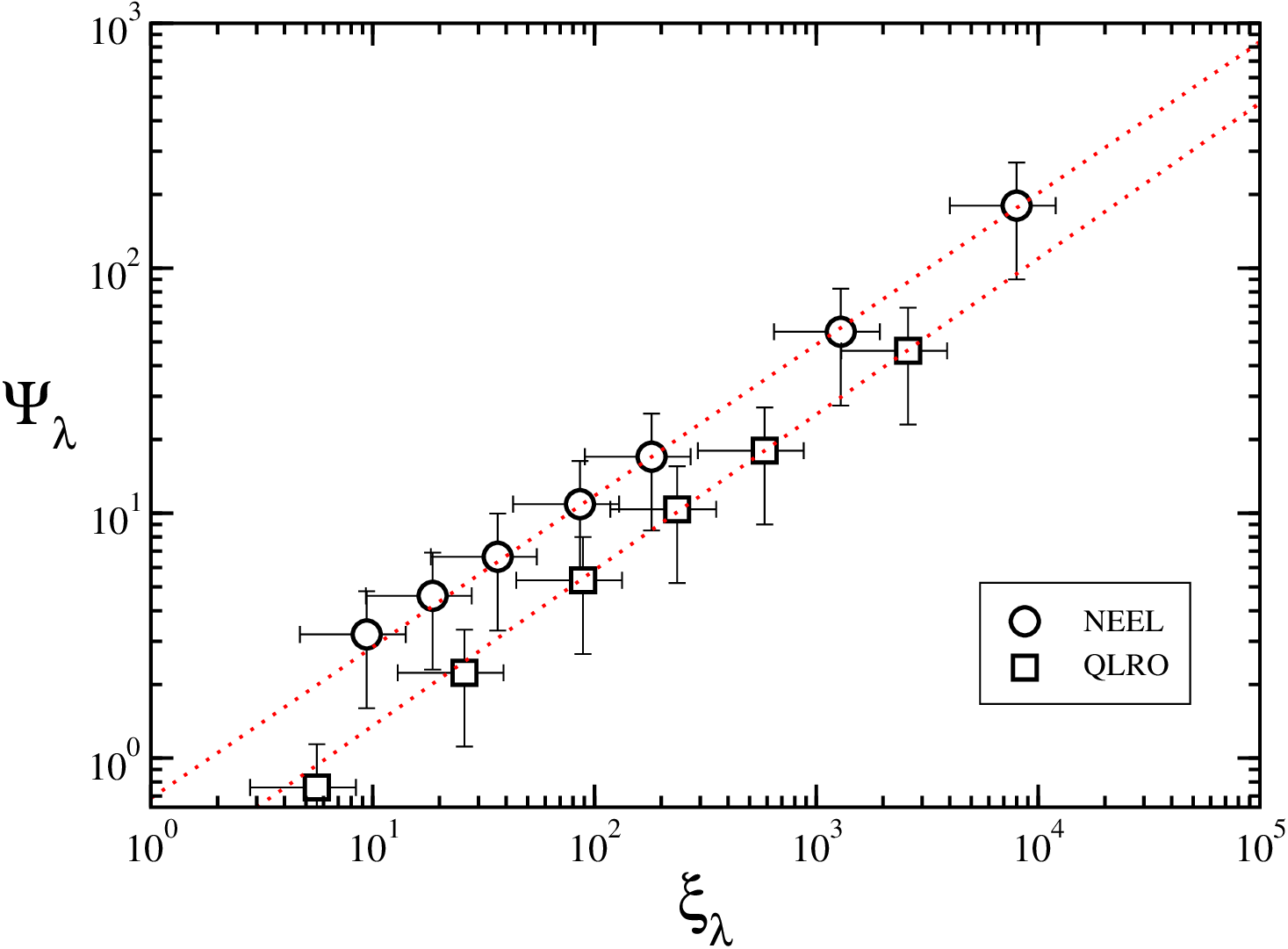}
\caption{Log-log plot of the scaling parameters $\Psi_{\lambda}$ and $\xi_{\lambda}$
  obtained from the collapses shown in Fig.~\ref{fig:cL}(b) for
  $\alpha=2.1$ in both phases: N\'eel ({\Large$\circ$}) and QLRO
  ($\square$). Within error bars (explicitly shown on the plot), data
  are fitted by power-laws (dotted lines) of the form
  $\xi_{\lambda}^{0.617}$ for the N\'eel regime and
  $\xi_{\lambda}^{0.636}$ for the QLRO.}
\label{fig:PsiXi}
\ec
\end{figure}

In the ordered phase, according to Eq.~(\ref{maf2}), we expect for the
AF order parameter
\be
m_{\rm{AF}}\propto \xi_{\lambda}^{\frac{1-z-\eta}{2}}\propto
(\lambda-\lambda_c)^{\frac{\nu(1-z-\eta)}{2}}.
\ee
This implies the usual {\it hyperscaling} relation involving the critical
exponent $\beta$ governing the onset of the order parameter
\be
m_{\rm{AF}}\propto (\lambda-\lambda_c)^{\beta},
\ee
which must therefore satisfy:
\be
2\beta=\nu(z+\eta-1),
\ee
{\em i.e.} the usual hyperscaling relation in $d=1$.

\subsubsection{Analytical estimate of the exponent $\eta$}
Following the same philosophy as in  the mean field argument given in
section~\ref{sec:heur}, we can calculate the expectation value of
the long range perturbation
\be
J(r,r')\langle {\vec S}_{r}\cdot {\vec
  S}_{r'}\rangle\sim \frac{1}{|r-r'|^{\alpha+z+\eta-1}},
\ee
at the QCP, with some unknown critical exponents
$z$ and $\eta$. The finite size correction to the free energy density
now scales like $L^{2-\alpha-z-\eta}$. The
singular part of the free energy at some non 
trivial QCP is expected to scale like $L^{-1-z}$ for a finite
size system. Thus the two corrections will scale in a similar way if 
\be
\label{MFeta}
\eta=3-\alpha.
\ee
The same condition is also obtained by 
demanding that the long range interaction be invariant under the RG
transformation involving a scale factor $s$: 
\be
\vec n(\tau ,x) \to s^{-(z-1+\eta )/2}\vec n(\tau /s^z,x/s).
\ee
(The rescaling factor of $s^{-(z-1+\eta )/2}$ for $\vec n$ implies the equal
time correlation exponent of $z-1+\eta$.)  
Rescaling $x$ and $\tau$ inside the integral Eq.~(\ref{Seff}) (which
represents the contribution of the long range term in the action), the
condition for invariance under this RG
transformation leads to $2+z-\alpha-(z-1+\eta)=0$, which gives the
expression (\ref{MFeta}) for $\eta$.
 As already shown in section~\ref{sec:SW},
Large-$N$ calculations also give  the same value for $\eta$~\cite{noteta}. 
Let us mention that the RG analysis, presented below in section~\ref{sec:RG}, also agree
with such an estimate, up to  order $(\alpha-2)^2$. 

\subsubsection{Numerical determination of the exponent $\eta$: scaling of
  the staggered susceptibility}
\begin{figure}[!ht]
\bc
\includegraphics[width=10cm,clip]{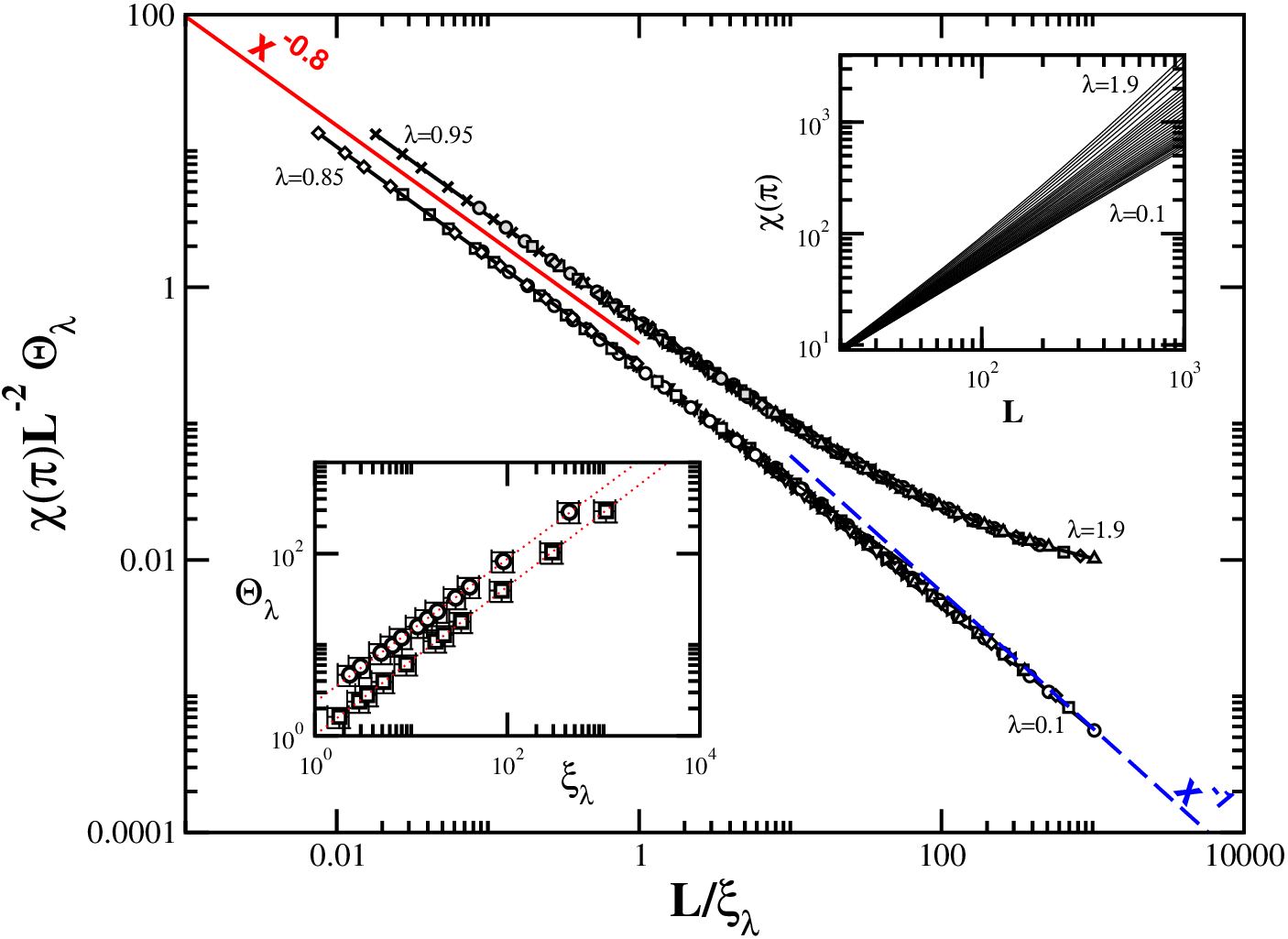}
\caption{QMC results for the $T=0$ staggered susceptibility
  $\chi(\pi)$ [Eq.~(\ref{def:stachi})] computed for $\alpha=2.2$ on
  systems up to $L=1024$ spins. The upper inset shows $\chi(\pi)$ vs $L$ for
  various $\lambda\in[0.1,1.9]$. The main plot shows the results of a
  data collapse onto two universal curves, after a rescaling of both
  $x$- and $y$-axis using two parameters $\xi_{\lambda}$ and
  $\Theta_{\lambda}$. 
Asymptotically, the LRO curve (top one with data
  for $0.95\le \lambda\le 1.9$) saturates
  towards a constant whereas the QLRO one (lower one with data for
  $0.1\le\lambda\le 0.85$) displays a
  $L^{-1}$ behavior, characteristic of $\eta=1$. Between them the
  separatrix shows the critical behavior around the transition at
  $\lambda_c\simeq 0.9$, decaying like $L^{-0.8\pm 0.01}$. The lower
  inset shows a log-log plot of the scaling parametrs $\Theta_{\lambda}$ and
  $\xi_{\lambda}$  used to achieve the collapses in both phases,
  N\'eel ({\Large$\circ$}) and QLRO ($\square$).
  Within error bars, data
  can be fitted for the entire range by power-laws (dotted lines) of
  the form $\xi_{\lambda}^{0.785}$ for the N\'eel regime and
  $\xi_{\lambda}^{0.815}$ for the QLRO.}
\label{fig:chi}
\ec
\end{figure}

The $T=0$ staggered susceptibility, defined on a finite ring of size
$L$ by
\be
\label{def:stachi}
\chi(\pi)=\frac{1}{L}\sum_{ij}(-1)^{|i-j|}\int_{0}^{\infty}\langle{\vec{S}}_{i}(0)\cdot{\vec{S}}_{j}(\tau)\rangle
d\tau,
\ee
obeys the standard finite size scaling at the QCP:
\be
\chi(\pi)\propto L^{2-\eta}.
\ee
Also we know, for instance from SW calculation (see
appendix~\ref{app:SW}), that in a N\'eel ordered state the staggered
susceptibility will scale quadratically with the size
$L$. On the other hand, in the QLRO characterized by $\eta=1$, we rather expect a linear
scaling of $\chi(\pi)$ with $L$. Consequently, there are
three distinct regimes for the staggered susceptibility:
\be
\label{eq:chi}
\chi(\pi)\times L^{-2}\sim\left\{
\begin{array}{lr}
{\rm{constant}}&{\rm{~if~}} \lambda > \lambda_c {\rm{~~~~(NEEL)}}\\
L^{-1} &{\rm{~if~}} \lambda < \lambda_c {\rm{~~~~(QLRO)}}\\
L^{-\eta} &{\rm{~if~}} \lambda = \lambda_c {\rm{~~~~(QCP).}}
\end{array}
\right.
\ee
We  use the same scaling procedure as  for the
correlation functions, to obtain data collapses onto two different
curves, as  illustrated in Fig.~\ref{fig:chi} for 
$\alpha=2.2$.  
Indeed $\chi(\pi)$,
  computed with QMC on chains of up to $L=1000$  
sites, displays clearly a crossover phenomenon on both sides of the
transition, also characterized by a crossover length scale which is
directly proportional to the one previously extracted in the analysis
of the correlation functions. Note also that such an analysis provides
a second physical observable way to locate the QCP: in fact, the analysis of $C(L)$
and $\chi(\pi)$ both agree (within the error bars) on the value of
$\lambda_c$.
Moreover, we expect the scaling hypothesis to be
valid if $\Theta_{\lambda}\sim \xi_{\lambda}^{\eta}$.
This is actually the case, as illustrated in the lower inset of Fig.~\ref{fig:chi}
where we find $\eta=0.8\pm 0.015$.
\begin{figure}[!ht]
\bc
\includegraphics[width=8cm,clip]{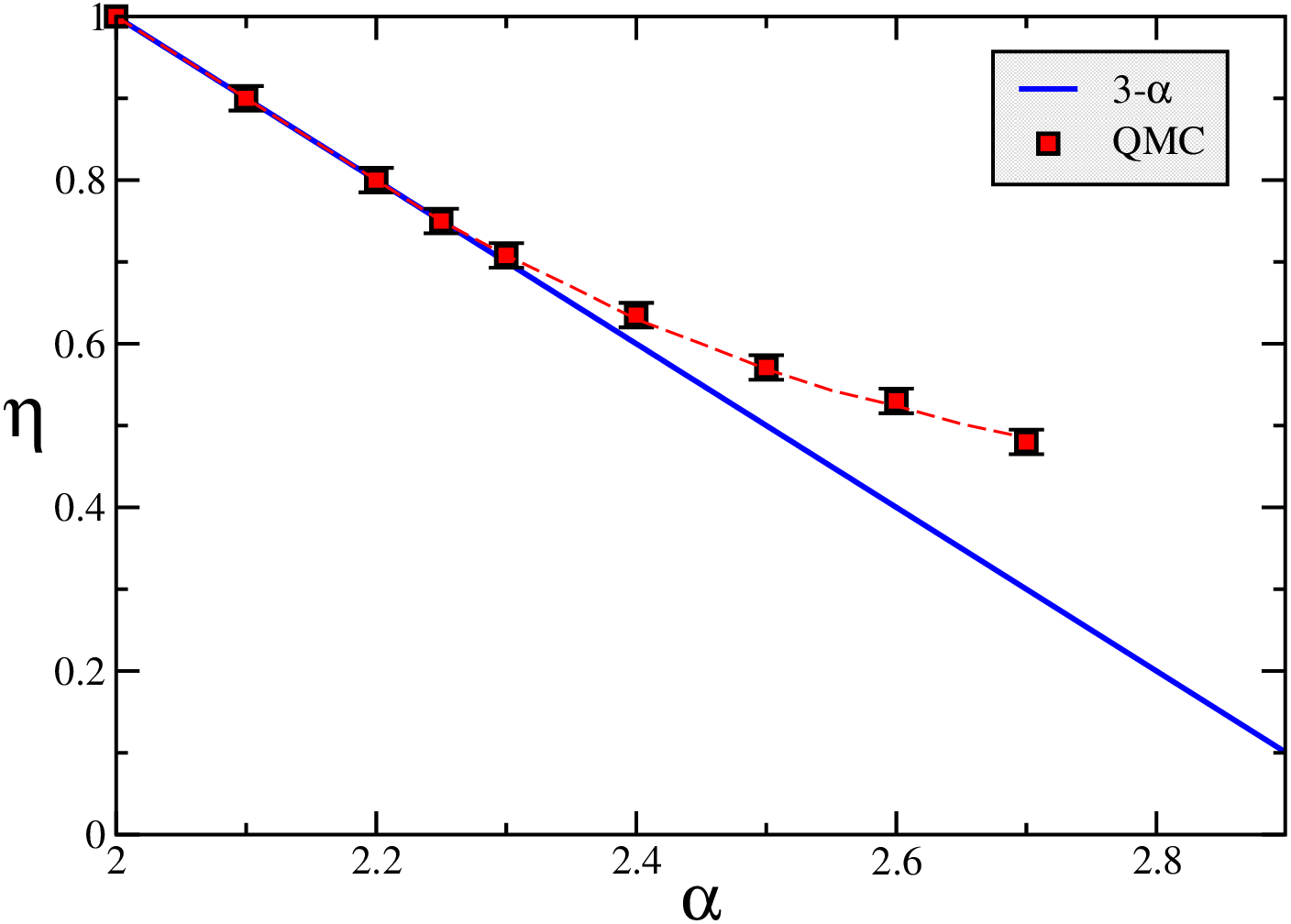}
\caption{Numerical estimate for the critical exponent $\eta$ along the
  critical line obtained using the critical behavior of $\chi(\pi)$
  computed with QMC (red
  symbols). The red dashed line is a guide to the eyes. The blue full
  line is the analytical estimate $\eta=3-\alpha$.}
\label{fig:eta}
\ec
\end{figure}
We can also obtain the quantum critical exponent
$\eta$ from the separatrix between the N\'eel and QLRO
regimes (see Fig.~\ref{fig:chi}), which is expected to decay as 
$L^{-\eta}$. For $\alpha=2.2$, we find $\eta=0.8\pm 0.01$.
We have repeated this computation of $\chi(\pi)$ for several other values of $\alpha
\in [2.1, 2.7]$ to calculate the corresponding $\eta(\alpha)$. The results are
plotted in Fig.~\ref{fig:eta}, and compared to the  previously
discussed estimate
$\eta=3-\alpha$. It is very remarkable to see 
how this rough estimate reproduces quite well the actual value. Only
for $\alpha>2.3$ a deviation starts to appear.}
\subsubsection{Dynamical exponent $z<1$}
The dynamical critical exponent $z$, involved for instance in the
critical decay of the correlation function Eq.~(\ref{eq:QCP}),
can be evaluated from the spin-spin correlation function at
the QCP. From the fit of the separatrix between the two data
collapses (see for instance Fig.~\ref{fig:cL} where for $\alpha=2.1$
we estimated $\eta+z-1=0.63\pm 0.03$) we obtain an estimate for
$\eta+z-1$. Then, using the estimates of $\eta$, determined separately with the
staggered susceptibility, we obtain a numerical evaluation of $z$.
Results are shown in Fig.~\ref{fig:z} for $2\le\alpha\le2.7$. For
$\alpha=2$, the QCP at $\lambda=0$ displays the critical behavior of
the short range model, with $\eta=z=1$. Surprisingly, when moving
from $\alpha=2$ along the transition line, $z$ becomes very rapidly
$<1$ and, within the error bars, seems to saturate around a value
$\sim 0.75$. It is actually natural to expect $z\neq 1$ since the long
range interaction breaks Lorentz invariance. However, unlike for the
estimate of $\eta$, the dynamical
exponent obtained within the large-$N$ expansion $z=(\alpha-1)/2$ does
not agree with the QMC results. As we discuss in the next section,
using a ``RG improved'' perturbation theory,  $z$ is
found to be $<1$ but does not display such a big reduction.
\begin{figure}[!ht]
\bc
\includegraphics[width=8cm,clip]{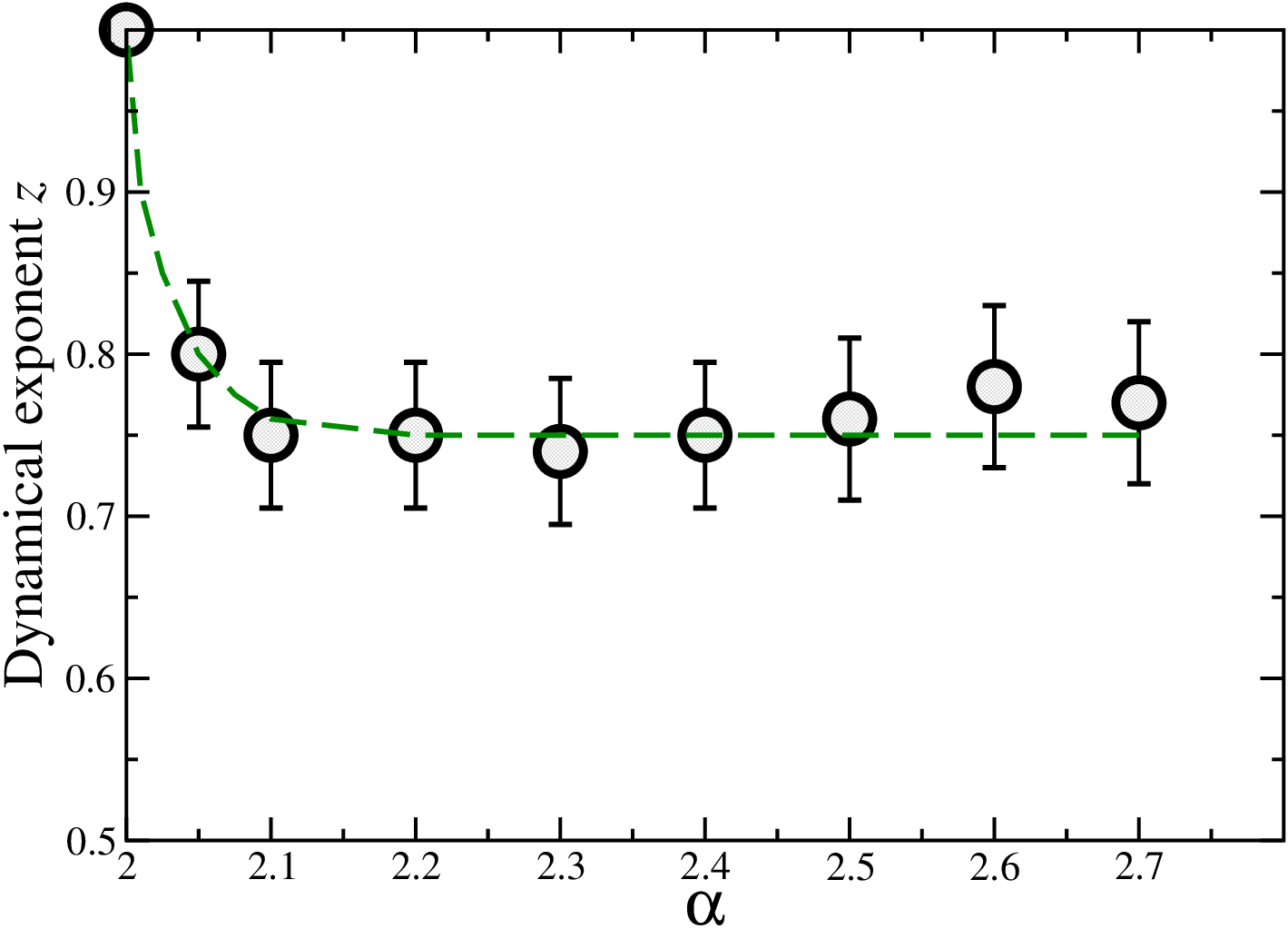}
\caption{Numerical estimates for the dynamical critical exponent $z$
  along the critical line obtained using the critical behavior of the
  correlation function [Eq.~(\ref{eq:QCP})] computed with QMC,
  and the numerical estimate of $\eta$ (see Fig.~\ref{fig:eta}). The
  numerical results (open circles) suffer from large error bars, as
  shown on the plot. The green dashed line is a guide to the eyes.}
\label{fig:z}
\ec
\end{figure}
\section{Field Theory/Renormalization Group Results}
\label{sec:RG}
The low energy, continuum limit, imaginary
time action takes the form: \be S[\vec n]=S_0[\vec n]-g\int d\tau
dx \phi (\tau ,x) -\lambda a^{\alpha -2}\int {d\tau dx dy\over
|x-y|^{\alpha}}\vec n(\tau ,x)\cdot \vec n(\tau ,y).
\label{Seff2}\ee
$\vec n(\tau ,x)$ is the antiferromagnetic order parameter field 
defined by the continuum limit expression of Eq. (\ref{cont_lim}).  
Here $S_0$ is the action for a free massless relativistic boson, in
terms of which $\vec n$ may be represented in a non-linear way.
Equivalently, we may regard it as the action of the $k=1$
Wess-Zumino-Witten non-linear $\sigma$ model. The field $\phi$ is
defined as: \be \phi = 2\pi\vec J_L\cdot \vec
J_R,\ee and is normalized so: \be <\phi (z)\phi (0)>={3\over
16\pi^2|z|^4}.\ee The corresponding coupling constant, $g$ has a bare value
of order 1 for the short range AF chain and is
marginally irrelevant. It is responsible for various logarithmic
corrections such as the one in the correlation function of
Eq. (\ref{corr}). Note that the
dimensionless coupling constant for the long range interaction
$\lambda$ is only proportional to the one used before for the lattice
microscopic model in Eq.~(\ref{QLRO}), and $a$
is a short distance cut-off with dimensions of length. As already
noted in  section~\ref{sec:heur},
since $\vec n$ has a scaling dimension of 1/2 from Eq. (\ref{ncorr}),
$\lambda$ is irrelevant for $\alpha >2$, relevant for $\alpha <2$, and
marginal for $\alpha =2$. Also note that $\lambda >0$ corresponds to
non-frustrating interactions which favor the N\'eel state where
$\langle n^z\rangle\neq 0$. Our strategy is to study this model when $0<\alpha -
2\ll 1$ and $0<\lambda \ll 1$ using perturbation theory in $g$ and
$\lambda$. Since $g$ renormalizes to $0$ at large length scales, when
$\lambda =0$ (i.e. in the short range model) this can give useful
results, for small bare $\lambda$, 
even when the bare value of $g$ is O(1).
 We will find that an interplay between the local marginal
coupling constant $g$ and the irrelevant non-local coupling constant
$\lambda$ governs the critical behavior in this regime.

We now consider the low energy effective field theory for the long
range model in Eq. (\ref{Seff2}), in the limit of small $g$ and
$\lambda$, using RG methods. When $\lambda =0$, the
RG equations reduce to the standard ones for
the short range model.  These take the form: \be {dg\over d\ln a}
=-g^2 -(1/2)g^3+\ldots .\label{RGg}\ee Here we define our RG
transformation by increasing the short distance cut-off $a$.  The bare
value of $g$ is positive for any 
non-frustrated 
short range model and is typically O(1).  The basin of
attraction of the $g=0$ fixed point is known to extend to such large
bare values of $g$ so that $g=0$ is the universal stable fixed point
for short range models. The flow of $g$ towards zero at long length
scales is controlled by the quadratic term in the $\beta$-function of
Eq. (\ref{RGg}), giving: \be g(a)\to {1\over \ln
(a/a_0)},\ee where $a_0$ is the original cut-off and $a$ is a larger
value obtained from integrating out modes with wave-lengths between
$a_0$ and $a$. This logarithmically slow flow of $g(a)$ to zero is
responsible for logarithmic corrections to the correlation function
and other properties of the short-range models.  A linear term in the
$\beta$-function for $\lambda$ follows immediately from the factor of
$a^{\alpha -2}$ in Eq. (\ref{Seff2}) which in turn is a consequence of
the fact that $n$ has scaling dimension $1/2$: \be {d\lambda \over
d\ln a}= (2-\alpha )\lambda +\ldots \label{RGl0}\ee So, ignoring the
effects of $g$, $\lambda$ grows larger at long length scales for
$\alpha <2$ but smaller for $\alpha >2$.  Long range interactions are
irrelevant for $\alpha >2$. However, it is necessary to consider
higher order terms in the $\beta$-functions for both $g$ and $\lambda$
to understand the phase diagram, even at $\alpha \approx 2$.

To calculate additional terms in the $\beta$-functions, we define our
ultra-violet cut-off by forbidding any 2 points in space-imaginary
time from getting closer than $a$, in a perturbative calculation of
the partition function (or long distance Green's functions). In
particular, this means that the long range term in the action is
cut-off as: \be S_{\lambda}=-\lambda a^{\alpha -2}\int_{|x-y|>a}
{d\tau dx dy\over |x-y|^{\alpha}}\vec n(\tau ,x)\cdot \vec n(\tau ,y).
\label{Sl}\ee
When the cut-off is increased from $a_0$ to $a=a_0+\delta a$, there is
an additional change in $S$ of first order in $\delta a$, which comes
from the change in the integration region: \be \delta S = -\lambda
a_0^{\alpha -2}\int dx d\tau \left[\int_{-a}^{-a_0} + \int_{a_0}^{a}
\right] {du\over |u|^{\alpha}}\vec n(\tau ,x)\cdot \vec n(\tau ,x+u)
.\label{deltaS}\ee Since both factors of $\vec n$ are very close
together, we may use the operator product expansion.  This follows
from the 3-point Green's function: \be <n^a(z_1)\phi
(z_2)n^b(z_3)>={1\over 8\pi}{|z_{13}|\over
|z_{12}|^2|z_{23}|^2}.\ee This implies the operator product expansion
(OPE): \be n^a(z)n^b(0)\to {\delta^{ab}2\pi \over 3}|z|\phi (0) +
\ldots \label{OPEnn}\ee 
Using this in Eq. (\ref{deltaS}), gives: \be \delta S =
-\lambda a_0^{\alpha -2}\int d\tau dx \phi (\tau ,x)
4\pi \int_{a_0}^a{du\over |u|^{\alpha -1}} \approx -{\delta a\over
a}4\pi \lambda \int d\tau dx \phi (\tau ,x).\ee This corresponds to
a renormalization of $g$: \be \delta g = 4\pi {\delta a\over
a}\lambda \ee and hence to another term in the $\beta$-function for
$g$: \be {dg\over d\ln a} = 4\pi \lambda -g^2
+\ldots \label{RGp}\ee There is one more term in the RG equations that
is important at small $\alpha -2$, corresponding to a correction to
$\lambda$ of order $\lambda g$. This can be calculated from the OPE:
\be \phi (z)n^a(0)\to {1\over 8\pi |z|^2}n^a(0) + \ldots ,\ee
giving: \be d\lambda /d\ln a =-(\alpha -2)\lambda +(1/2)\lambda g + \ldots
\label{RGl}\ee
The RG equations, Eq. (\ref{RGp}) and (\ref{RGl}) have an unstable
fixed point for $\alpha >2$, at: \bea g_c&\approx&2
(\alpha -2)\nonumber \\ \lambda_c &\approx& {1\over \pi}(\alpha
-2)^2.\eea For $\alpha <2$, a positive $\lambda$ always runs away to
large values as we lower the cut off (i.e. increase $a$),
corresponding to LRO.  On the other hand, for $\alpha
>2$, a small enough positive bare $\lambda$ flows to zero while a
larger bare value flows to large values (see Fig.~\ref{fig:RGflow}).
%%%%%%%%%%%%%%%%%%%%%%%%%%%%%%%%%%%%%%%%%
\begin{figure}[!ht]
\bc
\includegraphics[width=8cm,clip]{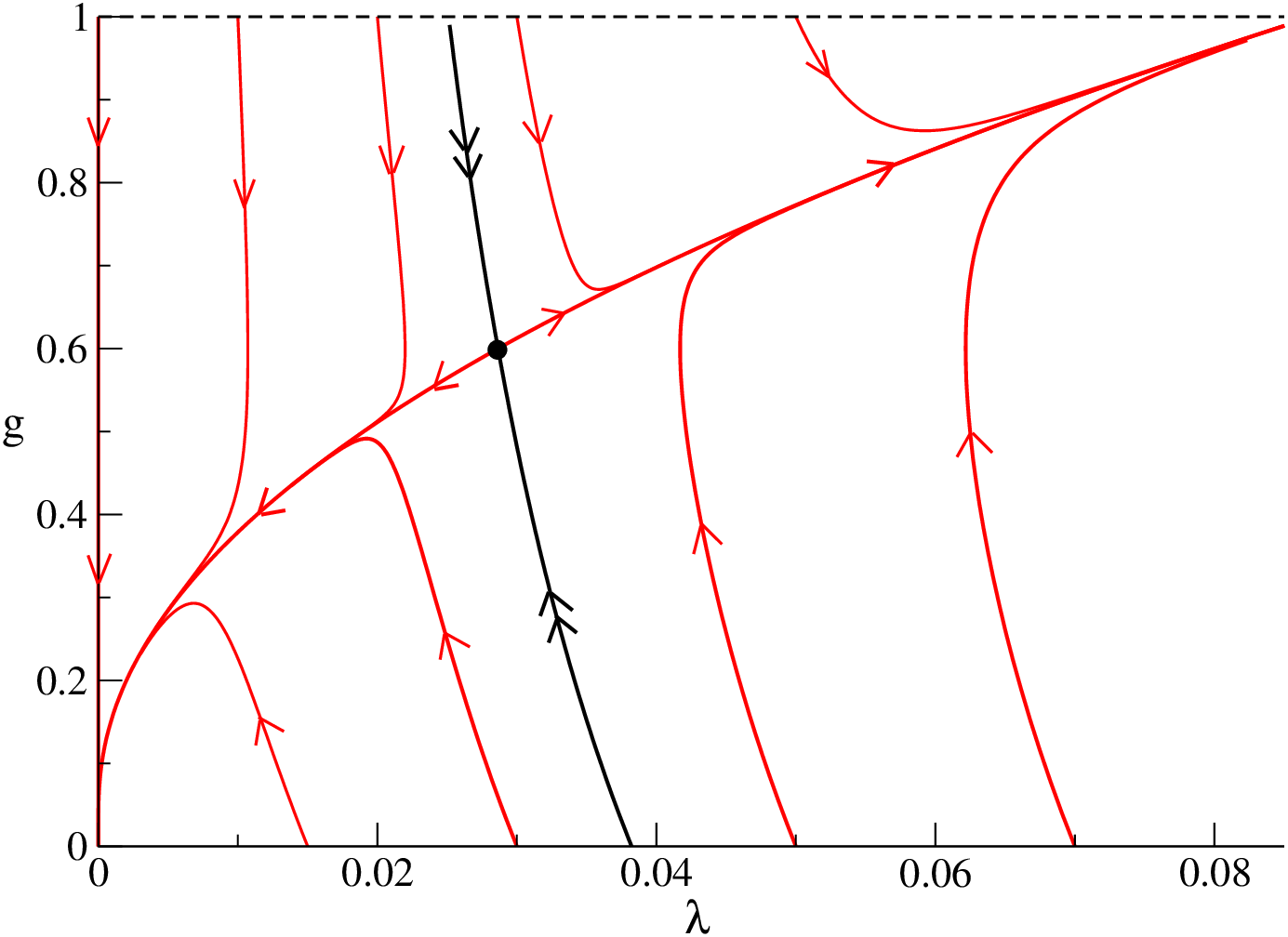}
\caption{Renormalization group flow of Eqs. (\ref{RGp}) and (\ref{RGl})
 in the case $\alpha=2.3$.
 The dotted line represents, schematically, the values of the bare
 couplings in the field theory as the parameter lambda in the lattice model
 is varied.  The unstable fixed point at $\lambda_c=0.0286$, $g_c=0.6$
separate the flow to the stable fixed point at $\lambda=g=0$ which
represents the quasi-long-range ordered phase and the flow to infinite
$\lambda$, $g$, which represents the long range ordered phase. The black
lines with double arrows denote the separatrixes between these two phases.
(The corrections to the flow equations are presumably significant for this
large a value of $\alpha -2$, but we graph this case for ease of
visualization.}
\label{fig:RGflow}
\ec
\end{figure}
%%%%%%%%%%%%%%%%%%%%%%%%%%%%%%%%%%%%%%%%%
These statements remain true
even when the bare value of $g$ is O(1) as we expect it to be in
general for a short range spin chain. For a small bare $\lambda$, $g$
initially renormalizes towards small values as it would in the short
range chain until eventually Eqs. (\ref{RGp}), (\ref{RGl}) becomes
valid. The stable $\lambda =g =0$ fixed points corresponds to the
standard QLRO phase of the short range spin chain.
The non-trivial unstable fixed point separates the ordered and quasi
long range ordered phases. 
Of course there are higher order terms in
both RG equations, but they do not invalidate our conclusions on the
location of the fixed point, for small enough $\alpha -2$. Both terms
on the right hand side of Eq. (\ref{RGp}) are $O[(\alpha -2)^2]$ at
the fixed point; any possible higher order terms such as $g^3$ or
$\lambda^4$ are at least of $O[(\alpha -2)^3]$.  Similarly, both
terms on the right hand side of Eq. (\ref{RGl}) are $O[(\alpha
-2)^3]$ at the fixed point; higher order terms are at least of
$O[(\alpha -2)^4]$.  To reach this conclusion it is important to
realize that there cannot be any terms in $d\lambda /d\ln a$ which
contain no factors of $\lambda$; a purely short range interaction
cannot generate a long range one although the reverse is not true.

Thus we appear to 
have a rare example of a non-trivial fixed point which can be
accessed perturbatively. (But see the discussion 
below of potential problems with this approach.) 
 We note that a similar expansion 
for long range classical spin models was introduced in \cite{Fisher72}. 
See also \cite{Sak,Bhatt}. 
Our quantum spin model, in the continuum limit, non-linear $\sigma$ 
model approximation, appears rather similar, in the 
imaginary time path integral formulation.  An important 
difference, however, is that our model has an action which 
is long range in the space direction but short-range in the time 
direction. Thus it corresponds to a classical model in 
2 space dimensions with short range interactions in one 
direction and long range interactions in the other. It is 
this asymmetry which leads to a dynamical critical exponent $z<1$.
Another important difference from a 2-dimensional Heisenberg 
model is the topological term in the short-range part of 
the action which is responsible for the quasi-long-range order.  
We remark that an integer-spin quantum Heisenberg chain 
with long range interactions could be expected to have 
identical critical behavior to a classical Heisenberg model 
in two dimensions with interactions which are long range in one dimension.  
We also remark that an xxz quantum spin chain with long 
range interactions could be expected to have the 
same critical behavior as a two-dimensional classical xy 
model with interactions which are long range in one dimension.

The phase boundary (or separatrix) can be found by determining the
line in the $g-\lambda$ plane which renormalizes to the critical
point.  Combining Eqs. (\ref{RGp}) and (\ref{RGl}) gives: \be
\int_{\lambda_0}^{\lambda_c}{d\lambda \over
\lambda}=(1/2)\int_{g_c}^{g_0}{dg(g-g_c)\over g^2-g_c^2(\lambda
/\lambda_c)},
\label{flow}\ee
where $\lambda = \lambda (g)$ is a function of $g$ along the RG flow
in the integral on the right hand side of Eq. (\ref{flow}). $g_0$ and
$\lambda_0$ are the values at arbitrary points on the separatrix.
Since $\lambda$ increases monotonically to the value $\lambda_c$ with
increasing $a$, we may obtain an upper and lower bound on the right
hand side by replacing $\lambda (g)$ by $\lambda_c$ and $0$
respectively inside the integral on the right hand side of
Eq. (\ref{flow}): \be (1/2)\ln (g_0/eg_c)+g_c/2g_0<\ln
(\lambda_c/\lambda_0)<(1/2)\ln [(g_0+g_c)/2g_c].\ee Now using the fact
that $g_c<<g_0$, this becomes: \be \sqrt{g_0/eg_c}<\lambda_c/\lambda_0
<\sqrt{g_0/2g_c},\ee that is: \be .637 (\alpha
-2)^{5/2}/\sqrt{g_0}<\lambda_0<.743 (\alpha
-2)^{5/2}/\sqrt{g_0}.\label{bound}\ee Thus on the separatrix
$\lambda_0$ is $O[(\alpha -2)^{5/2}]$.  [A numerical solution of the
RG equations indicates that $\lambda_0$ is very close to the lower
bound in Eq. (\ref{bound}).]  Assuming a bare $g$ of O(1), it is then
possible to predict the shape of the phase boundary in the $\lambda -
\alpha $ plane close to $\alpha =2$. There is an unknown
multiplicative factor relating the lattice coupling $\lambda$
to the continuum coupling $\lambda$. However, we can predict that \be
\label{eq:alphac}
\alpha_c(\lambda )\to 2+C\lambda^{2/5},
\ee for some unknown constant
factor, $C$, as $\alpha \to 2$. As mentioned above, we expect that
$2<\alpha _c(\lambda )<3$ for all $\lambda$ and all $S$.

While our QMC results also predict 
that $\alpha_c\to 2$ when $\lambda \to 0$, 
but otherwise Eq.~(\ref{eq:alphac}) does not agree well 
with the QMC result as shown in fig.~\ref{fig:zoom}.  
%%%%%%%%%%%%%%%%%%%%%%%%%%%%%%%%%%%%%%%%%
\begin{figure}[!ht]
\bc
\psfrag{A}{\tiny{$2+0.14\lambda^{2/5}$}}
\psfrag{B}{\tiny{$2+0.11\lambda^{2/5}$}}
\includegraphics[width=8cm,clip]{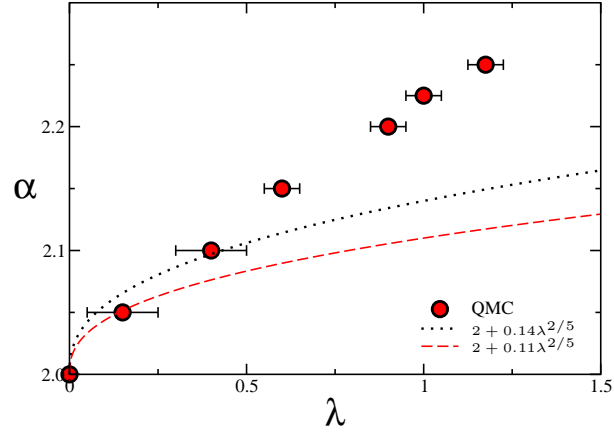}
\caption{QMC phase diagram at small $\lambda$ compared 
to the RG prediction of Eq. (\ref{eq:alphac}) with $C=0.14$ and $C=0.11$.}
\label{fig:zoom}
\ec
\end{figure}
%%%%%%%%%%%%%%%%%%%%%%%%%%%%%%%%%%%%%%%%%
It is interesting to note that lowest order SW theory and the large-$N$ approximation 
make the mean-field prediction 
that $\alpha_c \to 1$ as $\lambda \to 0$, in clear disagreement with
our RG and QMC results.

Linearizing the $\beta$-functions at the critical point gives: \be
{d\over d\ln a}\left(\begin{array}{c}\lambda -\lambda_c\\
g-g_c\end{array}\right)\approx \left(\begin{array}{cc}
0&(\alpha -2)^2/(2\pi)\\ 4\pi&-4(\alpha -2)\end{array}\right)
\left(\begin{array}{c}\lambda -\lambda_c\\ g-g_c\end{array}\right) \ee
This matrix has one positive (unstable) right eigenvalue,
$(\sqrt{6}-2)(\alpha -2)$, implying a crossover length scale: \be
\xi \propto |\lambda -\lambda_c|^{-\nu},\ee with a critical exponent:
\be \nu = {1\over (\sqrt{6}-2)(\alpha -2)},\ee which diverges as
$\alpha \to 2$.  The corresponding unstable direction is: \be
\lambda -\lambda_c={(\alpha -2)\over 2\pi(\sqrt{6}-2)}(g -g_c).\ee
For $\alpha =2$ exactly, LRO occurs for any $\lambda >0$ 
but behavior characteristic of the quasi long range ordered 
fixed point occurs out to a cross over length scale:
\be 
\xi \propto \exp [C/\lambda_0^{2/5}],
\label{eq:exp}
\ee
for a constant factor $C$. 

We may also determine the critical exponent, $\eta +z-1$, controlling
the equal-time correlation function at the non-trivial critical
exponent: \be <S_0^aS_j^b>\propto {\delta^{ab}\over |j|^{\eta
+z-1}},\label{eta}\ee when $\alpha -2<<1$.  Since $\lambda$ and $g$ are
both small at the critical point, for $\alpha -2<<1$, we may simply do
``RG improved'' perturbation theory.  That is we
calculate the correlation function to first order perturbation theory
in $g$, replace $g$ by $g_c$, and interpret the result as the
expression Eq. (\ref{eta}) with \be \eta + z = 2+ O(g_c).\ee [To
lowest order in $(\alpha -2)$ we only need consider $g$, not
$\lambda$, since $\lambda_c \propto g_c^2$.]  The Green's function, up
to first order in perturbation theory in $g$, the bare coupling, is:
\bea <n^a(z)n^b(0)>&=&{\delta^{ab}\over |z|}+g\int
d^2z'<n^a(z)n^b(0)\phi (z')>\nonumber \\ &=&{\delta^{ab}\over
|z|}+{g\over 2\sqrt{3}}\delta^{ab}\int d^2z'{|z|\over
|z'|^2|z-z'|^2}+\ldots \label{corr2}\eea The integral in
Eq. (\ref{corr2}) must be restricted to the region $|z'|>a$,
$|z-z'|>a$. For $|z|>>a$, the integral is dominated by the two
regions, $|z'|<<|z|$ and $|z'-z|<<|z|$, giving \bea
<n^a(z)n^b(0)>&\approx &{\delta^{ab}\over |z|}\left[1+{g\over
2\sqrt{3}}2\int {d^2z'\over |z'|^2}\right] \nonumber \\ &\approx
&{\delta^{ab}\over |z|}\left[ 1+{g\over 2\sqrt{3}}4\pi \ln
(|z|/a)\right] .\eea Now replacing $g$ by $g_c$, its fixed point value
we obtain: \be <n^a(z)n^b(0)>={\delta^{ab}\over |z|}\left\{1+(\alpha
-2)\ln (|z|/a)+O[(\alpha -2)^2]\right\} \ee Now, using the fact that
the correlation function should have a pure power-law form at the
fixed point, we may interpret this result as the leading term in the
expansion of: \be <n^a(z)n^b(0)>={\delta^{ab} A\over |z|^{1-(\alpha
-2)}}.\ee To this order in $(\alpha -2)$ we obtain the same exponent
for $z=ix$ or $z=\tau$, implying that the corrections to the dynamical
exponent, $z=1$, are higher order in $(\alpha -2)$.  Thus: \be \eta =
1-(\alpha -2) + O[(\alpha -2)^2].\ee 
This is the same value of $\eta$ found in the large $N$ approximation 
and by a simple scaling argument in sub-section (V-C-3). It 
agrees well with QMC results for $\alpha <2.3$ as shown in 
fig.~\ref{fig:eta}.
It is natural to expect that
$z\neq 1$ since the long range interaction breaks Lorentz invariance.
Since $\omega \propto |k|^{(\alpha -1)/2}$ in the ordered phase it is
natural to expect that $z<1$ at the critical point, as 
is found in the large $N$ approximation. However, clearly
$1-z$ must be at least of $O[(\alpha -2)^2]$ since the long range
coupling constant, $\lambda$ is of that order.  In fact, we suspect that
$1-z$ is even higher order than quadratic in $(\alpha -2)$.
This conclusion does not fit well with the QMC results 
for $z$, presented in fig.~\ref{fig:z}. There it was found 
(although with large error bars) that $z$ appears to have 
a nearly constant value, $z\approx .75$, for $\alpha \geq 2.1$. 
As $\alpha$ is further decreased $z$ appears to rise very rapidly 
towards $1$.  

So far, we have ignored another possible interaction:
\be S \to S -(g'\pi /3)\int d\tau dx (\vec J_L^2+\vec J_R^2).
\label{g'def}\ee
This interaction is, in fact, present for the short-range spin-chain 
with a large coefficient. Since the Hamiltonian for the 
$k=1$ WZW model can be written quadratic in currents, this 
``interaction'' term can be regarded as simply shifting the 
velocity, which we have so far set equal to $1$, to:
\be v \to 1-g'/2.\ee
The RG equations, for the short range model, including $g'$ take
the form, to cubic order:
\bea
{dg\over d\ln a}&=& -g^2-(1/2)g(g^2+g'^2)\label{RGg'1} \\
{dg'\over d\ln a}&=& (3/4)g^3.\label{RGg'2}\eea
Starting with $g$, $g'>0$ and O(1), these equations 
predict that $g\to 0$ and $g'$ flows to a value of O(1). 
The large value of $g'$ at the fixed point can simply 
be interpreted as a large renormalization of the velocity, 
provided that $g'<2$. In fact, this is what happens, 
for example in the Hubbard model at half-filling.  The 
spin velocity is reduced by the Hubbard interactions. 
An alternative approach is to adjust $v$ to the 
correct value and drop $g'$ completely from the RG 
equations. In fact, Eq. (\ref{RGg'2}) depends strongly 
on the renormalization and cut-off scheme. With 
a Lorentz invariant cut-off and renormalization procedure, 
the non Lorentz-invariant term, proportional to $g'$ will 
not be generated under the RG if it is initially absent. 
Breaking of Lorentz invariance in this problem at
low energies, just means shifting the velocity.  If 
we work directly with the exact velocity, then it is 
apparently permissible to set $g'=0$ and use 
a Lorentz invariant renormalization procedure so that 
$g'$ remains zero under renormalization. In fact, 
using this procedure in Eq. (\ref{RGg'1}) 
leads to various predictions of 
logarithmic corrections which are in good agreement 
with Bethe ansatz and numerical results for the S=1/2 chain. 
If we set $g'$ equal to some arbitrary non-zero value 
in Eq. (\ref{RGg'1}) the coefficients and powers of 
log corrections would change, resulting in worse 
agreements with numerical results. Thus, this 
procedure of ignoring $g'$ seems to be a valid and 
useful one. 

We now consider the interplay of the long range coupling 
constant, $\lambda$, with the non-Lorentz invariant 
local coupling, $g'$. The needed OPE's can be 
obtained from the general conformal field theory 
result for the 3-point Green's function of 
the energy momentum operator with a primary field of left-dimension $1/4$:
\be <T(z)n^a(z_1)n^b(z_2)>={1\over 2\pi}
\sum_{i=1}^2
 \left[{1/4\over (z-z_i)^2}+{1\over z-z_i}{\partial \over \partial z_i}
\right]{\delta^{ab}\over |z_1-z_2|}.\ee
Here $T=(2\pi /3)\vec J_L^2$, is the left-moving part of 
the Hamiltonian. This gives:
\be 
<T(z)n^a(z_1)n^b(z_2)>={1\over8\pi}{\delta^{ab}(z_1-z_2)^2 
\over |z_1-z_2|(z-z_1)^2(z-z_2)^2}.\ee
Also using the 2-point function of $T$:
\be <T(z_1)T(z_2))>={1\over 2(2\pi)^2(z_1-z_2)^4},\ee
we can deduce the OPE:
\be n^a(z)n^b(0)\to {\delta^{ab}\pi z^2\over |z|}T(0).\ee
Now consider the case where the separation is in the space direction, 
$z=ix$
\be n^a(x)n^b(0)\to  \delta^{ab}\pi |x|[(2/3)\phi (0)
-T(0)-\bar T(0)]+\ldots .\ee
Here we have included the identical OPE coefficient for 
$\bar T \equiv (2\pi /3)\vec J_R^2$ and also the coefficient 
deduced earlier, in Eq. (\ref{OPEnn}).
Using the same cut-off and RG transformation procedure as above, 
this implies a term in the $\beta$-function:
\be {dg'\over d\ln a}=-6\pi \lambda .\ee
This drives $g'$ towards negative values, corresponding to 
increasing the velocity. If we continue to use 
our previous RG transformation, we find that the 
$d\lambda /d\ln a$ {\it does not} pick up an term 
$\propto \lambda g'$.  The difference from the 
non-zero $\lambda g$ term in Eq. (\ref{RG3}) arises from 
the fact that the OPE is now:
\be   T(z)n^a(0)\to {1\over 8\pi z^2}n^a(0).\ee
The RG transformation gives:
\be \delta \lambda \propto \int {d^2z\over z^2},\ee
where, as before, the integral is over a circular shell 
with radius between $a$ and $a+\delta a$. This integral 
vanishes by rotational invariance. Thus the 
complete set of RG equations to low order is:
\bea {d\lambda \over d\ln a}&=&-(\alpha -2)\lambda +\lambda g/2\nonumber \\
{dg\over d\ln a}&=& 4\pi \lambda -g^2-(g+g')g^2/2\nonumber \\
{dg'\over d\ln a}&=&-6\pi \lambda .\label{RG3}\eea
These equations have an unstable fixed point at:
\bea \lambda_c&=&0\nonumber \\
g_c&=&2(\alpha -2)\nonumber \\
g'_c&=&-2-2(\alpha -2)\approx -2.\eea
In this approximation, we obtain the same prediction for $\eta 
\approx 1-(\alpha -2)$, as before and still get $z\approx 1$. 
There is a shift in the velocity of O(1). 

The large value of $g'$ at the fixed point makes the predictions 
of this RG analysis intrinsically suspect. As in the short-range 
case, we might agree to set $v$ equal to its renormalized 
value and then drop $g'$ from the RG equations. This leads 
to the same predictions about the value of $\lambda$ on 
the separatrix, and $\nu$ as obtained above. 

However, there are some worrisome features of this RG analysis 
which arise from the long range interaction. We expect 
a dynamical exponent $z<1$ 
at the critical point. It then does not make sense to use 
a rotationally invariant (i.e. Lorentz invariant) RG 
transformation.  We would then get back a $g^3$ term 
in $dg'/d\ln a$ and, very importantly, a $\lambda g'$ term 
in $d\lambda /d\ln a$. We would then generally find that 
$g$ is not small, O($\alpha -2$), at the fixed point. 
$\lambda$ would also not be small at the fixed point. 
In this case we would lose all perturbative control 
over the critical behavior even for $\alpha$ only 
slightly larger than $2$. In this case, the unstable critical 
point would not be close to the QLRO fixed point, for 
$\alpha$ close to $2$. One possibility is that the 
effects associated with $z<1$ can be ignored 
to lowest non-trivial order  $\alpha -2$. Then our 
use of a Lorentz invariant RG transformation 
may be justified.  In this case, 
the fixed point really is close to the QLRO critical point 
for $\alpha$ sightly larger than 2 and our predictions 
for $\eta$ and $\nu$ are correct in this limit. 

The QMC results seem to give at least 
partial confirmation of the validity of an RG 
approach based on the  $\alpha -2$ expansion.  Most 
importantly, $\alpha_c\to 2$ as $\lambda \to 0$ and 
the critical exponents $\eta$ and $z$ appear to approach their values 
in the quasi long range ordered phase ($\eta =z=1$) in this limit, 
with $\nu$ diverging. Furthermore, 
excellent agreement with the prediction for $\eta$ was obtained, 
over a rather large range of $\alpha$ (up to $2.2$), as shown 
in fig.~\ref{fig:eta}. We were not 
able to obtain accurate estimates of $\nu$ from QMC to 
test the RG prediction. On 
the other hand, $z$ showed rather surprising behavior, in fig.~\ref{fig:z}, 
dropping rapidly from $1$ to about $.75$ as $\alpha$ is increased 
from $2$ to $2.1$.  This suggests that the asympotic, small 
$\alpha -2$ behavior may only occur for very small values of 
$\alpha -2<<.1$. Numerical difficulties preclude obtaining 
QMC data in this region. Furthermore, the phase boundary
as determined by QMC, $\alpha_c(\lambda )$ 
could not be fit well to the RG prediction of Eq.~(\ref{eq:alphac}), 
except possibly at very small $\alpha_c$ (where we have no data) 
as seen in fig.~\ref{fig:zoom}. This could be interpreted 
as meaning that our RG approach based on an $\alpha -2$ expansion 
is correct in principle but is only valid in practice for 
extremely small values of $\alpha -2$.  (The good agreement 
for $\eta$ then appears fortuitous.) Alternatively, 
the discrepancies may indicate a problem with our RG approach, 
perhaps resulting from our cavalier treatment of the 
non-Lorentz invariant interaction in Eq.~(\ref{g'def}).

\section{Conclusions}
We have studied long range non-frustrating $S=1/2$ antiferromagnetic chains
using spin-wave theory, large-$N$ approximation, quantum Monte Carlo and 
analytic renormalization group methods based on an expansion 
in $\alpha -2$. All methods predict to a line of critical 
points in the $\lambda$-$\alpha$ plane with continuously varying 
critical exponents. This critical line separates phases with true
N\'eel long range order and quasi long range order. 
Quantum Monte Carlo and renormalization group methods 
indicate that this critical line terminates at $\lambda =0$, $\alpha
=2$ and suggest that, 
along the critical line as $\alpha \to 2^+$, 
 $\eta \approx 3-\alpha$, while $\nu$ diverges and $z\to 1^-$. 
\begin{acknowledgements}
We would like to thank J. Cardy for very helpful discussions.  
The research of N.L., I.A. and M.B. was supported by NSERC of Canada. 
The research of I.A. was supported by the Canadian Institute for 
Advanced Research. The numerical simulations were carried out on the
WestGrid network, funded in part by the Canada Foundation for
Innovation.

\end{acknowledgements}
\appendix
\section{Calculation of finite size corrections from SW:
  contributions from the $k=0$ and finite $k$ modes.}
\label{app:SW}

\subsection{General method}

Let ${\cal H}(h)={\cal H}-h\hat{O}$, 
where $\hat{O}$ is an operator  and $h$ is a  field. If we denote
by $|h\rangle$ the GS of ${\cal H}(h)$, it follows that the
GS energy is $E_{GS}(h) = \langle h | {\cal H}-h {\hat O}|
h\rangle$. Since $\langle h 
| h\rangle =1$, it is straightforward to show (in direct analogy to
Feynman's theorem) that:
\begin{equation}
\label{M1}
\langle \hat{O} \rangle= \langle h | {\hat O} |h\rangle = -
\frac{\partial E_{GS}(h)}{\partial h}  
\end{equation}

In general we need  the expectation values of
 various operators ${\hat O}$ in the
 unperturbed GS, i.e. in the limit $h \rightarrow 0$. It
 follows that all we have to do is to compute the change in the ground-state
 energy, due to the perturbation $-h\hat{O}$, to first order in $h$.

In the remainder of this Appendix, the  Hamiltonian ${\cal H}$ is that
of Eq. (\ref{QLRO}). We are interested in finite-size chains with an
even number of sites $L$, and periodic boundary conditions.

\subsection{Staggered susceptibility}

Let $\hat{O}=\sum_{i}^{}(-1)^i S^z_i$. Then, according to
Eq. (\ref{M1}),  the staggered magnetization at $T=0$ is:
$$
M_{\pi}= \langle \sum_{i}^{}(-1)^i S^z_i \rangle = -\left.{d
  E_{GS}\over dh}\right|_{h\rightarrow 0}
$$
and therefore the staggered susceptibility is:
$$
\chi(\pi) = {1 \over L} \left.{d M_{\pi} \over
  dh}\right|_{h\rightarrow 0} = - {1\over L} \left. {d^2
	E_{GS} \over d h^2}\right|_{h\rightarrow 0}.
$$

\subsubsection{The $k=0$ contribution}
We Fourier transform the spin operators, $\vec{S}_{2n}= 2/L
\sum_{k}^{} \exp{(ik(2na))} 
\vec{S}^{e}_k$, $\vec{S}_{2n+1}= 2/L \sum_{k}^{} \exp{(ik(2n+1)a))}
\vec{S}^{o}_k$,  and collect only the $k=0$
components. Let us denote $\vec{S}_{1} = \vec{S}^{e}_{k=0}=\sum_{n}^{}\vec{S}_{2n}
$ and  $\vec{S}_{2} = \vec{S}^{o}_{k=0}=\sum_{n}^{}\vec{S}_{2n+1}$ the
total spins of the two magnetic sublattices, of $L/2$ spins
each. Since we are in a N\'eel 
ordered state, $\vec{S}_1$ and $\vec{S}_2$ are spins of total
magnitude $S_L=LS/2=L/4$ for spins 
$S=1/2$. Then, up to some constants that do not depend on $h$:
\be
{\cal H}_{k=0}(h) = {\cal H}_{k=0} -{\cal H}_1
\ee
where
\be
{\cal H}_{k=0} = j \vec{S}_1\cdot \vec{S}_2,
\ee
\be
{\cal H}_1=h \left( S_1^z-S_2^z\right)
\ee
and 
\be
j = {2 \over L} \left[1+\lambda  \sum_{n\ge 1} {1 \over
	(2n+1)^\alpha} \right]= {2J_{eff}\over L} 
\ee

We need to do perturbation theory to second order in $h$, to find the
staggered susceptibility. The
ground-state of ${\cal H}_{k=0}$ is the state:
\be
|0\rangle = |S_T=0, M_T=0, S_L, S_L\rangle 
= {1 \over \sqrt{2S_L+1}}
\sum_{m=-S_L}^{S_L}(-1)^m |m, -m\rangle
\ee
where $\vec{S}_T = \vec{S}_1 + \vec{S}_2$. 
The perturbation  links this only to other states with $M_T=0$ (see below). Let
us denote \be
|S_T\rangle = |S_T,0,S_L,S_L\rangle
\ee
where $S_T=0,1,...,2S_L$. With this notation, the
second-order correction to the ground-state energy is:
\be
\Delta E_{\rm GS}^{(2)} = \sum_{n=1}^{2S_L} \frac{|\langle n | {\cal H}_1| 0\rangle|^2}{E_0-E_n}
\ee

However,
\be
{\cal H}_1 |0\rangle = {2h\over \sqrt{2S_L+1}} \sum_{m=-S_L}^{S_L} (-1)^m m
|m, -m\rangle
\ee
Interestingly enough, one can show that:
\be
|1\rangle = \alpha_1 \sum_{m=-S_L}^{S_L} (-1)^m m
|m, -m\rangle
\ee
where the normalization constant is:
\be
\alpha_1=\sqrt{3\over S_L(S_L+1)(2S_L+1)}.
\ee
It follows:
\be
\langle 1 |{\cal H}_1|0\rangle = {2h\over \sqrt{2S_L+1}\alpha_1}
\ee
and  $\langle n |{\cal H}_1|0\rangle = 0$, $\forall n \ge 2$. By
direct calculation, we find:
\be
E_0 = -jS_L(S_L+1); E_1 = j - jS_L(S_L+1)
\ee
and therefore:
\be
\Delta E_{\rm GS}^{(2)} = \frac{|\langle 1 | {\cal H}_1|
  0\rangle|^2}{E_0-E_1} = -{4h^2\over (2S_L+1)\alpha_1^2 j}
= -{4h^2\over 3j}S_L(S_L+1)
\ee
Since $j=2J_{eff}/L$, $S_L=LS/2$, we find:
\be
\Delta E_{\rm GS}^{(2)} = -{h^2\over 3J_{eff}}L^2S\left({L\over 2}S+1\right).
\ee
As a result, the contribution of the $k=0$ modes to the staggered susceptibility
is:
\be
\chi_{k=0}(\pi) = - {1\over L} \left.{d^2 E_{GS}\over d h^2}\right|_{h=0}
= {2 \over 3}{LS\over J_{eff}}\left({L\over 2}S+1\right)\sim  L^2
\ee

\subsubsection{The contribution of finite $k$ modes}

Within the spin-wave approximation, the contribution of the $k\ne 0$
  modes to the ground-state energy of ${\cal H} - h \hat{O}$ is the
  zero-mode energy:
\be
E_{GS} \sim 
  \sum_{k\ne 0 }^{}\omega_k=\sum_{k}^{}\sqrt{(\gamma-f(k)+h)^2-g^2(k)}
\ee
Here, the spin-wave dispersion is changed by the addition of the
perturbation $-h \sum_{i}^{}(-1)^i S^z_i$. The second derivative of
$E_{GS}$ with respect to $h$ can now be calculated trivially, and in
the limit $h\rightarrow 0$ we find:
\be
\chi_{k \ne 0}(\pi)\sim {1\over L}\sum_{k}{g^2(k)\over \omega_k^{3}} 
\ee
where $g(k)$ was defined before Eq. (\ref{hsw}).
As $k \rightarrow 0$, $g(k)\rightarrow \gamma=const$, $\omega_k
\sim k^{\alpha-1\over 2}$, and  therefore
\be
\chi_{k \ne 0}(\pi)\sim{1\over L}^{1-3{\alpha-1\over2}}\sim L^{3\alpha-5\over 2}
\ee
If  $\alpha <3$, this is a smaller power than the $L^2$ contribution obtained from the
$k=0$ mode. It follows that within the N\'eel ordered state, the
staggered susceptibility scales like $L^2$ (at least within the
spin-wave approximation).

\subsection{Transverse correlation function}

We now choose 
\be
\hat{O} = \sum_{i}^{}S^+_{i+n} S^-_{i}
\ee
so that its expectation values $\langle \hat{O}\rangle$ is the
transverse contribution to C(n). 
For simplicity, we  assume $n$ to be  even (calculations can be done similarly
for odd $n$). Since this is not a Hermitian operator, let:
\be
\hat{O}_A = \sum_{i}^{}\left(S^+_{i+n} S^-_{i}+S^+_{i-n} S^-_{i}\right)
\ee
\be
\hat{O}_B = i\sum_{i}^{}\left(S^+_{i+n} S^-_{i}-S^+_{i-n} S^-_{i}\right)
\ee
Both these operators are hermitian. According to Eq. (\ref{M1}) and
using invariance to translations, we
then have
\be
L \langle S^+_{n} S^-_{0}\rangle = \langle \sum_{i}^{}S^+_{i+n} S^-_{i}\rangle
= \langle \hat{O}_A\rangle - i \langle \hat{O}_B \rangle \rightarrow
\langle S^+_{n} S^-_{0}\rangle = -{1 \over L}
\left[{dE_{GS}^{(A)}\over dh} -i {dE_{GS}^{(B)}\over dh} \right]
\ee
where $E_{GS}^{(A/B)}$ are the ground-states energies in the presence
of perturbations $-h  \hat{O}_{A/B}$. 

\subsubsection{The contribution of finite $k$ modes}

After a Fourier transform, we use the Holstein-Primakoff
representation 
for all $k \ne 0$ modes. Keeping only quadratic terms, we find:
\be
\hat{O}_A = 4S \sum_{k}^{}\cos(nka) \left(b_k^{\dagger} b_k + a_k^{\dagger} a_k \right)
\ee
\be
\hat{O}_B = 4S \sum_{k}^{}\sin(nka) \left(b_k^{\dagger} b_k - a_k^{\dagger} a_k \right)
\ee

After adding this to the unperturbed Hamiltonian (in the SW approximation)
and diagonalizing, we find the ground-state energies to be:
\be
E_{GS}^{(A)} = JS \sum_{k \ne 0 }^{}\omega_{k, A}  +4sh \sum_{k\ne 0}^{} \cos(nka)
\ee
\be
E_{GS}^{(B)} = JS \sum_{k \ne 0 }^{}\omega_{k, B} - 4sh \sum_{k\ne 0}^{} \sin(nka)
\ee
where 
\be
\omega_{k, A} = \sqrt{[\gamma - f(k) -{4h\over J} \cos(nka)]^2-[g(k)]^2}
\ee
\be
\omega_{k, B} = \sqrt{[\gamma - f(k) +{4h\over J} \sin(nka)]^2-[g(k)]^2}
\ee

After taking the first derivatives and setting $h=0$, we  find the
$k\ne 0$ modes'
contribution to the transverse correlation to be:
\be
\langle S^+_n S^-_0\rangle = {2S\over L} \sum_{k \ne 0}^{}\frac{(\gamma
  - f(k)) e^{ikna}}{\omega_k} -{2S\over L} \sum_{k \ne 0}^{} e^{ikna}
\ee
In the limit $k \rightarrow 0$, $f(k) \rightarrow 0$,
$\omega_k\rightarrow k^{\alpha-1\over 2}$ and therefore:
\be
\langle S^+_n S^-_0\rangle= a_1 L^{\alpha-3\over 2}( 1 + ...) +2S/L
\ee
The second term is the second sum ( $\sum_{k \ne 0}^{} e^{ikna} = 
\delta_{n,0}L/2 - 1 =-1$, since $n >0$).
For $\alpha > 2$, the $L^{\alpha-3\over 2}$ term is dominant. 

\subsubsection{The $k=0$ mode}

Keeping  the full $k=0$ contributions, we find
\be
{\cal H}_{k=0}(h) = j \vec{S}_1\cdot \vec{S}_2 - {2h \over L} \left(S_1^+S_1^- +
S_2^+ S_2^-\right)
\ee
The notation has been introduced in the previous section. The
ground-state $|0\rangle$ of ${\cal H}_{k=0}$ is known (see section on
staggered susceptibility), so the first order contribution to $E_{GS}(h)$
can be evaluated directly:
\be
L \langle S^+_{n} S^-_0\rangle = -\left.{d E_{GS} \over
  dh}\right|_{h=0} \rightarrow 
\langle S^+_{n} S^-_0\rangle = {2 \over L^2} \langle 0 | S_1^+S_1^- +
S_2^+ S_2^-|0\rangle
\ee
The calculation is trivial, and we find:
\be
\langle S^+_{n} S^-_0\rangle = {2 S^2\over 3}\left[1 + {2\over LS}
  \dots \right] \sim  {1 \over L}
\ee
It follows that for this correlation function, the finite $k$ modes
give the dominant $L$ dependence, which is $L^{\alpha-3\over 2}$.

This calculation can be repeated for the parallel contribution to the
correlation, $\langle S^z_{n} S^z_0\rangle$. The $L$ dependence
remains the same, so we conclude that in the N\'eel state and within
SW approximation, $C(L) \sim L^{\alpha-3\over 2}$.

\section{Large-$N$ calculation}
\label{app:LN}
Considering the order parameter $\vec \phi$ is a $N$-component unit vector
field
\be |\vec \phi (\tau ,x)|^2=1,\label{constraint}\ee
the action can be written like
\be
S={N\over 2g}\int d\tau dx [(\partial \vec \phi /\partial \tau)^2
+(\partial \vec \phi /\partial x)^2]
-\lambda N
\int d\tau dx dy \vec \phi (\tau ,x)\cdot \vec  \phi (\tau ,y)/|x-y|^{\alpha}.
\ee
We have set $v=1$. $g\propto 1/s$ is a coupling constant, not related to what we 
called $g$ in other sections. $g$ and $\lambda$ are scaled 
by $N$ in order to have a smooth large-$N$ limit. Inside a path integral, we may integrate over all 
fields $\vec \phi (\tau ,x)$, without the constraint of Eq. (\ref{constraint})
provided that we introduce a Lagrange multiplier field, $\sigma (\tau ,x)$:
\be S \to S+{iN\over 2g}\int d\tau dx \sigma (\vec \phi^2-1).\ee
The action is now quadratic in  unconstrained fields, so that we may, in 
principle, do the Gaussian integration over $\vec \phi$. For this purpose 
it is convenient to write the long range term in $\omega$-$k$ space
using:

\be 
\int_a^\infty dx{e^{ikx}\over |x|^\alpha}\approx {2\over (\alpha -1)}
\left(a^{-(\alpha -1)}-|k|^{\alpha -1}\Gamma (2-\alpha )\sin [(2-\alpha  )\pi /2]\right).
\label{int} 
\ee

Here $a$ is a short distance cut-off and this equation is valid for
$|k|a\ll 1$. 
$\Gamma$ is Euler's Gamma function.  Note that the prefactor blows up, $\propto 1/(\alpha -1)$ 
as $\alpha \to 1$. 
Note also that for $1<\alpha <2$, both $\Gamma (2-\alpha )$ and
 $\sin [\pi (2-\alpha  )/2]$ are $>0$, so that the second term in Eq. (\ref{int}) is $>0$.  
As $\alpha \to 2$, $\Gamma (2-\alpha )\to 1/(2-\alpha)$ so that $\Gamma (2-\alpha )
\sin [\pi (2-\alpha )/2]\to \pi /2$. For $2<\alpha <3$, both $\Gamma (2-\alpha )$ 
and $\sin [\pi (2-\alpha )/2]<0$ so that their product is positive, blowing up as $\alpha \to 3$. 

The first term in Eq. (\ref{int}) can be eliminated by shifting $\sigma$ by a (imaginary) constant. 
Thus we may write:
\be 
S={NV\over 2g}\int {d\omega dk\over (2\pi )^2} \vec \phi (\omega ,k)\cdot \vec \phi (-\omega ,-k)
[\omega^2+k^2+C\lambda |k|^{\alpha -1}] +{iN\over 2g}\int d\tau dx \sigma (\vec \phi^2-1).
\ee
Here
\be C\equiv {4\over \alpha -1}\Gamma (2-\alpha )\sin [\pi (2-\alpha )/2]\ee
and $V$ is the space-time volume. 
Now we do the functional integral over $\vec \phi$. This gives an effective action for the
 field $\sigma$, which has an overall factor of $N$ in front of it, and no further dependence on $N$. 
At large $N$ the functional integral over $\sigma$ is dominated by a saddle point corresponding 
to a constant and purely imaginary value of $\sigma $. Assuming $\sigma$ is constant, 
this effective action is:
\be S_{eff}/V = {N\over 2}\left\{ \int {d\omega dk\over (2\pi )^2}\ln [\omega^2+k^2+C|k|^{\alpha -1}+i\sigma]
-{i\sigma \over g}\right\}\ee
The saddle point is found by looking for a stationary point of $S_{eff}$. Setting 
$i\sigma =m^2$ at the saddle point gives the self-consistent equation which 
determines $m^2$:
\bea
{1\over g}&=&\int {d\omega dk\over (2\pi )^2}{1\over \omega^2+k^2+C\lambda |k|^{\alpha -1}+m^2}\nonumber\\
&=&\int^\Lambda_{-\Lambda} {dk\over 4\pi }{1\over \sqrt{k^2+C\lambda
    |k|^{\alpha -1}+m^2}}
\eea
Here we have introduced an ultra-violet cut-off, $\Lambda$ of O($a^{-1}$). 
For any $\alpha$, $\lambda$ and $g$ for which this equation has a solution with $m^2>0$, 
the system is in the disordered phase, with a finite gap, $m$. Note that, for $\alpha >3$, 
the integral diverges as $k\to 0$ when $m=0$.  For small finite $m$ it behaves as
$\ln (\Lambda /m)$. Since this diverges as $m\to 0$, there will always be a solution for $m$, 
no matter how small is $g$. On the other hand, for $1<\alpha <3$, there will be a 
solution for $\lambda <\lambda_c$ only. At the critical value of $\lambda$, $m=0$, so 
$\lambda_c$ is determined by:
\be {1\over g}=\int^\Lambda_{-\Lambda} {dk\over 4\pi }{1\over \sqrt{k^2+C\lambda_c |k|^{\alpha -1}}}\ee
As $\alpha \to 1$, $C\to 4/(\alpha -1)$ and this becomes:
\be {1\over g}=\int^\Lambda_{-\Lambda} {dk\over 4\pi }{1\over \sqrt{k^2+4\lambda_c /(\alpha -1)}}\ee
This we see that $\lambda_c\propto (\alpha -1)$, as $\alpha \to 1$.  $\alpha =1$ is the critical 
value of $\alpha$, for which $\lambda_c\to 0$.  The behavior of $\lambda_c(\alpha )$ 
is qualitatively similar to what is obtained from SW theory, including the 
behavior near $\alpha =1$. Right at the critical point, the AF spin-correlation function
is determined by the effective action with $m=0$, and $\vec \phi$
treated as a 
non-interacting, free field, 
in the large $N$ approximation.  This implies a dispersion relation:
\be \omega =\sqrt{k^2+C\lambda |k|^{\alpha -1}},\ee 
and hence a dynamical exponent:
\be z=(\alpha -1)/2.\ee
This dispersion relation is the same as in SW theory, but there is no long range 
order at the critical point and hence no ambiguity in the value of $z$. The spin correlation function 
is given by:
\be <\phi^a(\tau ,x)\phi^b(0,0)>=\delta^{ab}{g\over N}\int {d\omega dk\over (2\pi )^2}
{e^{i(\omega \tau +kx)}\over \omega^2+k^2+C\lambda |k|^{\alpha -1}}.\ee
In particular, the equal-time correlation function is:
\be <\phi^a(0,x)\phi^b(0,0)>=\delta^{ab}{g\over N}\int {dk\over (4\pi )}
{e^{ikx}\over \sqrt{k^2+C\lambda |k|^{\alpha -1}}}.\ee
At large $x$, we may approximate this by dropping the $k^2$ term. It then follows from 
a rescaling that this decays as:
\be <\phi^a(0,x)\phi^b(0,0)>\propto {\delta^{ab}\over |x|^{(3-\alpha )/2}}.\ee
The standard definition of the critical exponent $\eta$ then implies:
\be z-1+\eta = (3-\alpha )/2,\ee
giving:
\be \eta =3-\alpha .\ee
Note that this is simply the behavior of the transverse correlation function in the 
ordered phase, according to SW theory. 

To calculate $\nu$, we need to calculate how $m$ vanishes with $\lambda_c-\lambda$ 
as $\lambda \to \lambda_c$, from below. It turns out that there are two 
different behaviors, depending on whether $1<\alpha \leq 5/3$ or $5/3\leq \alpha <3$.  A small change in $\lambda$ leads to a change in $1/g$ 
which is linear in $\delta \lambda$. 
Consider the effect of a small non-zero $m$ on the gap equation. For $1<\alpha \leq 5/3$, 
we may Taylor expand the gap equation:
\be 
{1\over g}\approx 
\int^\Lambda_{-\Lambda} {dk\over 4\pi }\left[{1\over \sqrt{k^2+C\lambda |k|^{\alpha -1}}}-(m^2/2){1\over [k^2+C\lambda |k|^{\alpha -1}]^{3/2}}\right]
\ee
(Note that this integral is finite at $k\to 0$ for $(3(\alpha -1)/2<1$ 
only, implying $\alpha <5/3$.)
Setting $\lambda =\lambda_c$, we see that:
\be 1/g-1/g_c \propto -m^2.\ee
Thus $m^2\propto -\delta \lambda$ in this case. Now consider 
the equal time correlation function for a small $m^2$.
\be <\phi^a(\tau ,x)\phi^b(0,0)>=\delta^{ab}{g\over N}\int { dk\over (4\pi )}
{e^{ikx}\over \sqrt{k^2+C\lambda |k|^{\alpha -1}+m^2}}.\ee
At large distances and small $m^2$ we should be able to drop the $k^2$ term 
and ignore the ultra-violet cut-off.  A rescaling of the $k$-integration 
variable then implies a correlation length:
\be \xi \propto m^{-2/(\alpha -1)}\propto (\delta \lambda )^{-1/(\alpha -1)}
\ee
and hence an exponent:
\be \nu = 1/(\alpha -1),\ \  (1<\alpha \leq 5/3 ).\ee

Now consider the case $5/3\leq \alpha <3$. Keeping a small non-zero 
$m^2$, we find:
\be {d\over dm^2}{\left( 1\over g\right) }={-1\over 2}
\int {dk\over 4\pi}{1\over [\lambda C|k|^{\alpha -1}+m^2]^{3/2}}.\ee
Since the integral is dominated by $|k|$ of $O(m^{2/(\alpha -1)}$, 
we have taken the cut-off to infinity and dropped the $k^2$ term. 
By scaling, we see that:
\be {d\over dm^2}{\left( 1\over g\right) }\propto m^{(5-3\alpha )/(\alpha -1)}
.\ee
Integrating with respect to $m^2$ gives:
\be { 1\over g}\approx {1\over g_c}-Am^{(3-\alpha )/(\alpha -1)},\ee
for a constant, $A$. Thus we see that:
\be m\propto (-\delta \lambda )^{(\alpha -1)/(3-\alpha )}.\ee
This gives:
\be \xi \propto m^{-2/(\alpha -1)}\propto (-\delta \lambda )^{-2/(3-\alpha )}.
\ee
Thus:
\be \nu = {2\over 3-\alpha},\ \  (5/3 \leq \alpha <3).\ee


\begin{thebibliography}{999}
\bibitem{note1} The proof for non-existence of LRO at $T=0$ for
  $\lambda =1$, if $\alpha
  >3$ \cite{Parreira97}, can be trivially extended and
  shown to hold for   all $\lambda \ne 1$.
\bibitem{note2} These equations are the analogs of equations (6) of
  Ref. \cite{Yusuf04}. We use $\gamma$ instead of $\alpha$ since
  in our work $\alpha$ defines the spatial variation of the exchange
  (the parameter called $\beta$ in Ref. \cite{Yusuf04}). The
  definition of $f(k)$ is corrected by a factor of $2$.
\bibitem{notexi}
A precise determination of
  the exponent $\nu$ is actually impossible to achieve
  because of large error bars for $\xi_{\lambda}$ due to some natural
  uncertainties in the data collapse procedure.
\bibitem{noteta} 
It is important to note that, up to  small corrections, this
estimate Eq.~(\ref{MFeta}) was
also found 3 decades ago by several authors for the related  $n$-vector
model with long range interaction~\cite{Fisher72,Brezin76,Ma73}.
\end{thebibliography}
\end{document}